\documentclass[sigconf, nonacm]{acmart}

\usepackage{tikz}
\usepackage{amsmath}
\usepackage{color}
\usepackage{graphicx}
\usepackage{colortbl}
\usepackage{xspace}
\usepackage{subfig}
\usepackage{multirow}
\usepackage{pifont}
\usepackage{algorithm}
\usepackage{algorithmicx}
\usepackage{enumitem}

\usepackage{amssymb}
\usepackage{microtype}
\usepackage[dvipsnames]{xcolor}
\usepackage{balance}
\usepackage{mathtools}

\newcommand{\cmark}{\ding{51}}%
\newcommand{\xmark}{\ding{55}}%

\newcommand{\old}[1]{}

\newcommand{\proposed}[0]{ZipFlow\xspace}
\newcommand{\pattA}[0]{Fully-Parallel\xspace}
\newcommand{\pattB}[0]{Group-Parallel\xspace}
\newcommand{\pattC}[0]{Non-Parallel\xspace}

\newcommand{\fig}[1]{Figure~\ref{#1}}
\newcommand{\sect}[1]{Section~\ref{#1}}
\newcommand{\tab}[1]{Table~\ref{#1}}

\newcommand{\tpch}[0]{TPC-H\xspace}

\begin{document}
\title[\proposed{}: a Compiler-based Framework to Unleash Compressed Data Movement for Modern GPUs]{\proposed{}: a Compiler-based Framework to Unleash \\ Compressed Data Movement for Modern GPUs }

\author{Gwangoo Yeo}
\affiliation{%
  \institution{KAIST}
  \city{Daejeon}
  \state{South Korea}
}

\author{Zhiyang Shen}
\affiliation{%
  \institution{Tsinghua University}
  \city{Beijing}
  \country{China}
}

\author{Wei Cui}
\affiliation{%
  \institution{Microsoft Research Asia}
  \city{Vancouver}
  \country{Canada}
}

\author{Matteo Interlandi}
\affiliation{%
  \institution{Microsoft Gray Systems Lab}
  \city{Los Angeles}
  \country{United States}
}

\author{Rathijit Sen}
\affiliation{%
  \institution{Microsoft Gray Systems Lab}
  \city{Redmond}
  \country{United States}
}

\author{Bailu Ding}
\affiliation{%
  \institution{Microsoft Research}
  \city{Redmond}
  \country{United States}
}

\author{Qi Chen}
\affiliation{%
  \institution{Microsoft Research Asia}
  \city{Vancouver}
  \country{Canada}
}

\author{Minsoo Rhu}
\affiliation{%
  \institution{KAIST}
  \city{Daejeon}
  \country{South Korea}
}

\begin{abstract}

In GPU-accelerated data analytics, the overhead of data transfer from CPU to GPU becomes a performance bottleneck when the data scales beyond GPU memory capacity due to the limited PCIe bandwidth. Data compression has come to rescue for reducing the amount of data transfer while taking advantage of the powerful GPU computation for decompression. To optimize the end-to-end query performance, however, the workflow of data compression, transfer, and decompression must be holistically designed based on the compression strategies and hardware characteristics to balance the I/O latency and computational overhead. In this work, we present \proposed, a compiler-based framework for optimizing compressed data transfer in GPU-accelerated data analytics. \proposed classifies compression algorithms into three distinct patterns
based on their inherent parallelism. For each pattern, \proposed employs generalized scheduling strategies to effectively exploit the computational power of GPUs across diverse architectures. 
Building on these patterns, \proposed delivers flexible, high-performance, and holistic optimization, which substantially advances end-to-end data transfer capabilities.
We evaluate the effectiveness of \proposed on industry-standard benchmark, \tpch.
Overall, \proposed achieves an average improvement of $2.08\times$ over the state-of-the-art GPU compression library (nvCOMP) and $3.14\times$ times speedup against CPU-based query processing engines (e.g., DuckDB).

\end{abstract}

\maketitle

\section{Introduction}

GPUs have become widely available with the rapid development of artificial intelligence, such as deep neural networks and large language models. Interestingly, given the powerful computation capability and high memory bandwidth of modern GPUs, they also exhibit great potential for analytical database workloads. Recent work has demonstrated the promise of leveraging GPUs to speed up analytical database systems, including efforts from both academia and industry~\cite{gpuaccelerationsqlanalytics, ocelat,crystal,tqp,sigmod22Hu, Lee21VLDB, blazingdb, heavydb, funke18sigmod}.

\begin{figure}[t!] \centering
\includegraphics[width=0.45\textwidth]{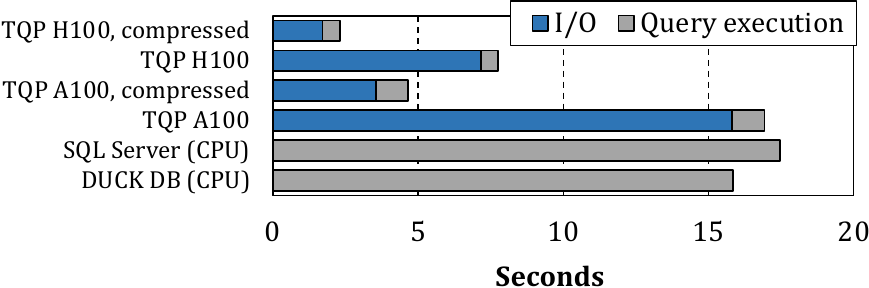}\\
    \caption{Latency breakdown of 22 TPC-H queries (SF=100) on CPU (AMD EPYC 7V12)~\cite{sqlserver, duckdb}, A100 (PCIe-4), and H100 (PCIe-5), {TQP~\cite{tqp} is used for query processing on GPU.
    } 
    }
    \vspace{-1.2em}
\label{fig:pcie_is_bottleneck}
\end{figure}

Despite such promising results, the constrained GPU memory capacity has been a critical limitation for GPU-accelerated database systems. For example, high-end GPUs such as NVIDIA H100\cite{h100} only have up to 80 GB of memory capacity, while analytical databases often scale to hundreds of GB or even several TBs. Although using multiple GPUs connected with high-performance interconnect (i.e., NVLink~\cite{nvlink}) can increase the memory capacity at high memory bandwidth, such hardware can be prohibitively expensive. As the database cannot be fully cached in the GPU memory, the data often needs to be transferred from CPU to GPU via the PCIe bus on demand during query execution
~\cite{pvldb23Cao, gaming, vldb10Fang, sigmod22shanbhag, yinandyang}.  
Unfortunately, the data transfer often becomes a performance bottleneck due to the slow PCIe bus. For example, \fig{fig:pcie_is_bottleneck} shows the query latency breakdown of the 22 \tpch queries on CPU-based and GPU-based database systems. Although the query execution on A100~\cite{a100} is more than $15\times$ faster than that on CPU, due to the slow data transfer from CPU to GPU, the speedup of the end-to-end query latency is much more moderate.

Fortunately, analytical database systems typically store data in a compressed format (e.g., Parquet~\cite{parquet}). This allows the data to be initially compressed on the CPU and later decompressed by the GPU before query processing. This design choice reduces the amount of data transferred and simultaneously leverages the computational power of the GPU.
While data compression has already become the status quo for analytical database systems, we observe that existing data compression techniques are designed for CPU-based database systems and do not fully exploit the unique performance characteristics of GPU-accelerated database systems.

First, GPUs offer massive computation that can unlock the power of more aggressive compression strategies for increased compression ratios to improve end-to-end query performance. Existing data compression techniques often limit the choices of compression algorithms and their nesting to strike a balance between the compression ratio and decompression overhead. For example, the widely used Parquet format only considers compute-efficient compression algorithms, such as run-length-encoding (RLE), delta encoding, dictionary encoding, and bit-packing, while more expensive compression algorithms, such as LZ77-based and entropy-based algorithms, can lead to much higher compression ratios on some workloads. Even state-of-the-art GPU-based compression libraries provide limited support for nested compression schemes. For example, nvCOMP~\cite{nvcomp} provides nesting only for specific combinations of RLE, delta encoding, and bit-packing.

Second, compression schemes must be tailored to the target GPU hardware to achieve optimal performance. While specialized algorithms that adapt to data characteristics can be highly effective~\cite{afroozeh2024alp, fsst}, mapping them respectively onto a wide variety of GPU architectures entails a substantial overhead for adapting utilization on each target platform. Existing GPU compression frameworks~\cite{nvcomp, sigmod22shanbhag, afroozeh2024accelerating} provide libraries of pre-built kernels only for conventional algorithms, while manually extending these frameworks to incorporate new or custom compression techniques is both cumbersome and inefficient.
In addition, these frameworks often lack the support for optimizing nested compression schemes, i.e., no kernel fusion for cross-stage optimization with pre-built kernels. 
Furthermore, achieving seamless kernel compatibility across heterogeneous GPU architectures, such as support for NVIDIA GPUs via CUDA~\cite{cuda} and AMD GPUs via ROCm~\cite{hipify}, continues to present a formidable challenge for existing frameworks.

Finally, optimizing end-to-end query performance requires holistic design of the data compression, transfer, and decompression workflow. This includes exploring the tradeoffs between compression ratio, decompression speed, and data transfer schedules to minimize overall latency.
While prior work has studied the performance of individual components~\cite{sigmod22shanbhag, btrblocks, corra, afroozeh2024accelerating} or examined designs with a limited set of algorithms~\cite{hippogriff,daemon25nicholson}, such localized decision-making can lead to suboptimal end-to-end performance.

To address these gaps, we propose \proposed, a flexible compiler-based compression framework for modern GPUs.
\proposed abstracts commodity compression algorithms into three parallel compute patterns. These patterns serve as building blocks, enabling flexible and generalized device scheduling of both existing and new custom algorithms.
Each pattern is easily optimized to exploit its parallelism across diverse GPU architectures with \proposed's matching compiler technique.
By holistically exploring the tradeoffs between I/O and computation overhead of the compressed data transfer workflow, \proposed delivers optimized end-to-end query performance.
Our evaluation on the industry-standard \tpch benchmark demonstrates the benefits of \proposed. Our custom compression methods provide superior compression ratios, with $1.85\times$ savings in I/O latency compared to a state-of-the-art compression GPU library~\cite{nvcomp}. Meanwhile, \proposed also provides a $3.26\times$ speedup in decompression with automated compiler optimizations.
This leads to an average speedup of $2.08\times$ in end-to-end query latency compared to the baseline. 

In summary, this work makes the following contributions:
\begin{itemize}[leftmargin=*]
    \item We abstract three patterns to encapsulate the core parallelism structures commonly found in (de)compression algorithms (Section~\ref{sect:design}). These patterns enable native optimization of existing and custom (de)compression algorithms on GPUs, which can also facilitate development of a broader set of GPU-friendly (de)compression algorithms in the future.

    \item For each pattern, we formulate a universal optimization space attuned to GPU execution geometries and the SIMT paradigm (\sect{sect:implementation}). Combined with fusion and pipelining, it holistically minimizes the end-to-end compressed data transfer overhead.

    \item We conduct a head-to-head comparison between \proposed and the state-of-the-art nvCOMP on identical algorithms across a wide range of data distributions (\sect{sect:pattern_experiment}), highlighting that our hardware-aware approach effectively
    outperforms the state-of-the-art GPU compression library (i.e., nvCOMP).
    
    \item We comprehensively evaluate \proposed on the industry-standard \tpch benchmark (Section~\ref{sect:query_level_experiment}), demonstrating the impact of joint optimization of the compressed data transfer workflow on the end-to-end query performance. Finally, we showcase how \proposed enables tailored hardware-aware execution by evaluating its performance on heterogeneous GPUs.
    
\end{itemize}

\section{Background}

\subsection{Existing Compression Algorithms}
\label{sect:compression_background}

\begin{figure}[t!] \centering
\includegraphics[width=0.48\textwidth]{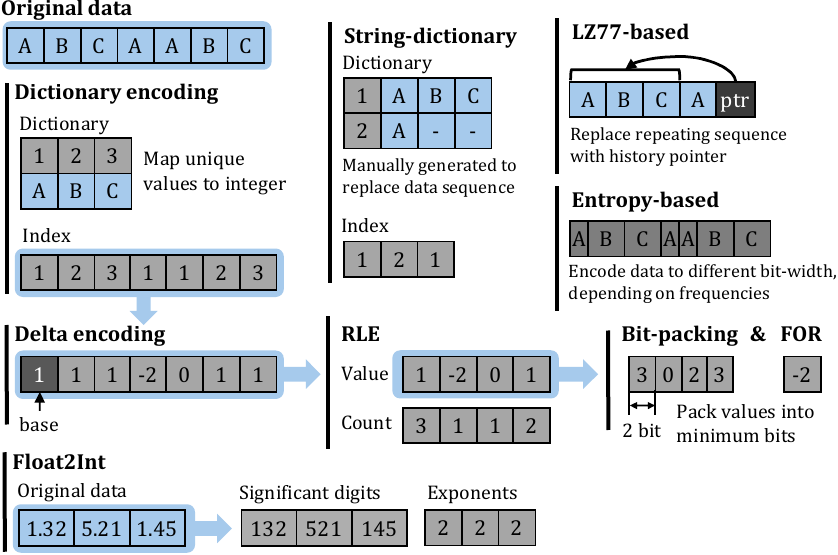}\\
	\caption{Compression algorithms of different families.}
    \vspace{-1.0em}
\label{fig:compression_bg}
\end{figure}

In this section, we present a summary of some of the widely used lossless compression algorithms and discuss the fundamental design principles behind them. We also illustrate an example diagram for each algorithm in \fig{fig:compression_bg}.

\noindent
{\bf Dictionary encoding} replaces data with a \textit{dictionary} and an \textit{index}. The \textit{dictionary} stores unique values from data, while the \textit{index} maps original data values to their corresponding entries in the dictionary.

\noindent
\textbf{Delta encoding} replaces each value with the difference (delta) between the current value and previous value, while storing the initial value as a base. Delta encoding alone does not provide compression, but it enables further compression with RLE or bit-packing.

\noindent
{\bf Run-Length Encoding (RLE)} minimizes consecutive repeating integer values by replacing them with a \textit{value} and a \textit{count}.
The \textit{value} stores the repeated value, while the \textit{count} records the number of repetitions. 

\noindent
{\bf Bit-packing and Frame of Reference (FOR)} pack integers into the minimum bit-width required for the given data.
The bit-width is calculated using minimum and maximum values, while the minimum value is stored separately as a FOR.

\noindent
{\bf Float2Int} separates significant digits and exponents from floating-point values, converting them into integers, enabling further compression. This approach is effective for compressing columns with limited decimal precisions. The core idea is shared in several prior works~\cite{btrblocks,vldb10Fang, afroozeh2024alp,galp}, though the name and details differ.

\noindent
\textbf{String-dictionary} de-duplicates repeated byte sequences by substituting them with dictionary indices, and the dictionary is constructed to store high-frequency byte sequences within the target dataset (from prior works ~\cite{fsst, gsst}).
The compression ratio depends on how well the generated dictionary de-duplicates the patterns.

\noindent
\textbf{Entropy-based algorithms} such as ANS~\cite{ans, ansgit} and Huffman encoding~\cite{huffman} compress the data by exploiting the skewness of data occurrences. Fewer bits are assigned to frequently occurring data to reduce overall data size.

\noindent
\textbf{LZ77-based algorithms} such as LZ4~\cite{lz4} and Snappy~\cite{snappy} de-duplicate repeated sequences by encoding them as pointers to previous data. These algorithms are referred to as “general-purpose” due to their broad compatibility. State-of-the-art algorithms such as Deflate~\cite{deflate} and zstd~\cite{zstd} combine LZ77 with entropy-based methods for superior compression ratios.

\subsection{Existing Approaches in Database Systems}
In modern database systems, achieving efficient data compression is often accomplished through the \emph{composition of several lightweight compression techniques}, such as run-length encoding (RLE), delta encoding, dictionary encoding, and bit-packing. This paradigm, frequently described as \textit{cascaded compression}, leverages the strengths of individual algorithms by orchestrating them in a sequence, tailored to the characteristics of the data at hand. Well-designed nested algorithms for individual data files show an effective compression ratio, while offering simple compute patterns~\cite{DammeHHL17}. 
{BtrBlocks~\cite{btrblocks} introduces an automated search mechanism to determine optimized nested algorithms, significantly improving compression ratios.}

Prominent columnar storage formats such as ORC~\cite{orc} and Parquet~\cite{parquet} have embraced this cascaded approach by implementing a variety of well-engineered nested compression schemes, such as combining RLE with dictionary encoding or integrating delta encoding with bit-packing. They support further compression by enabling general-purpose algorithms to be applied on top of predefined nesting algorithms.
The Fastlanes project~\cite{fastlanes} was introduced to enhance compressed columnar data formats by enabling increased opportunities for parallel decoding, 
providing several handcrafted GPU kernels, such as bit-packing, delta encoding~\cite{afroozeh2024accelerating}, and Float2Int~\cite{galp}, aiming to leverage the SIMD and SIMT processing capabilities of CPUs and GPUs.

\section{Design}
\label{sect:design}

\begin{figure}[t!] \centering
\includegraphics[width=0.44\textwidth]{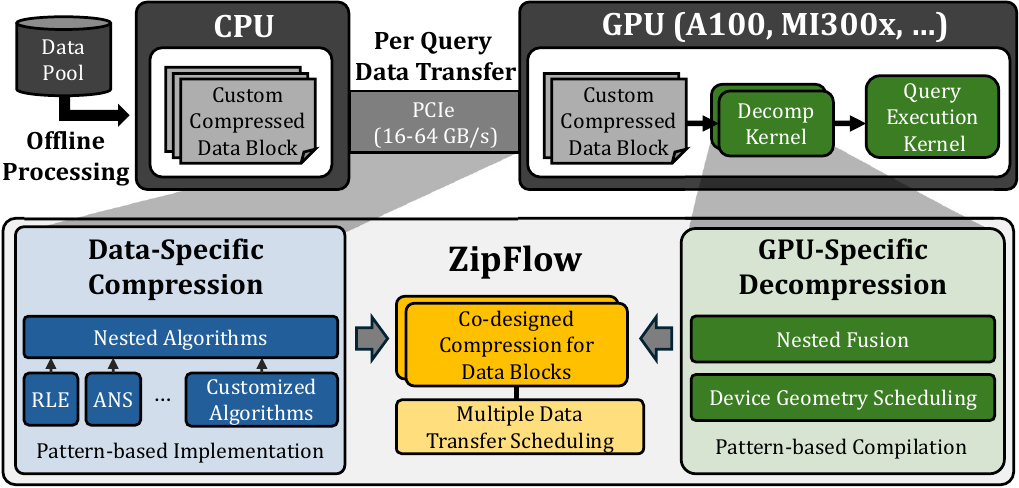}\\
	\caption{End-to-end execution pipeline from stored data to \proposed.}
\label{fig:data_movement_scenario}
\end{figure}

\begin{figure}[t!] \centering
\includegraphics[width=0.44\textwidth]{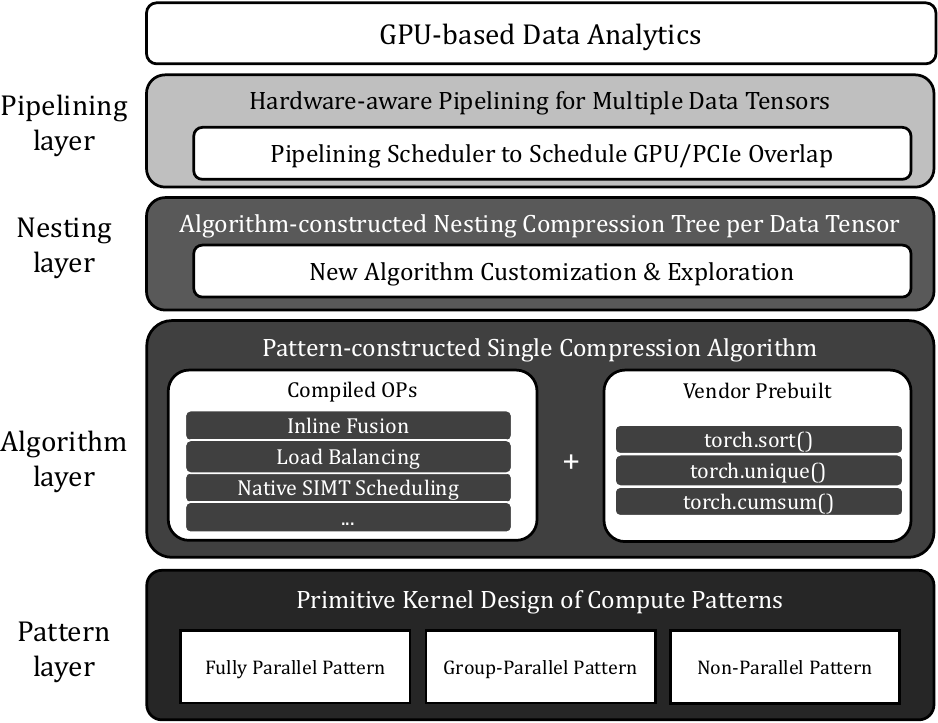}\\
    \caption{{\proposed Architecture Overview.}}
\label{fig:framework_system}
\end{figure}

We propose \proposed, a framework to accelerate compression-based data movement to GPUs.
\fig{fig:data_movement_scenario} presents a high-level overview of the end-to-end data movement pipeline as optimized by \proposed. Raw data in various formats (e.g., Numpy, Parquet, Text, Binary) are first compressed utilizing \proposed flexible nesting and customizable algorithms, with the resulting compressed representations stored in CPU memory. For online data transfer, \proposed facilitates the efficient movement of compressed data across the PCIe interconnect, followed by an ultra-fast, device-specific decompression phase. This end-to-end process is co-designed by \proposed to fully exploit the heterogeneity of modern GPUs, incorporating not only hardware-native resource management but also optimized overlap between PCIe data transfers and on-device decompression to maximize throughput and minimize latency.

\fig{fig:framework_system} introduces a four-layer design for \proposed.
Starting from the bottom, the \textbf{Pattern Layer} enables the design of primitive GPU operators based on three predefined compute patterns. Each designed operator corresponds to a GPU kernel and serves as a building block for (de)compression algorithms.
These GPU operators enable simple implementation of the target algorithm while exploiting the prevalent parallel patterns in (de)compression algorithms.
In the \textbf{Algorithm Layer}, GPU patterns are organized to form primitive compression algorithms such as RLE, dictionary encoding, bit-packing, or ANS.
Based on primitive algorithms from the Algorithm Layer, the \textbf{Nesting Layer} explores and customizes the nesting of algorithms applied to data columns (e.g. Dictionary encoding + ANS or RLE + Bit-packing), to achieve optimal end-to-end data movement latency.
Finally, at the top is the \textbf{Pipelining Layer}, where our framework schedules the pipelining of I/O operations and GPU decompression latency across multiple data chunks to minimize overall latency.

\begin{figure}[t!] \centering
\includegraphics[width=0.48\textwidth]{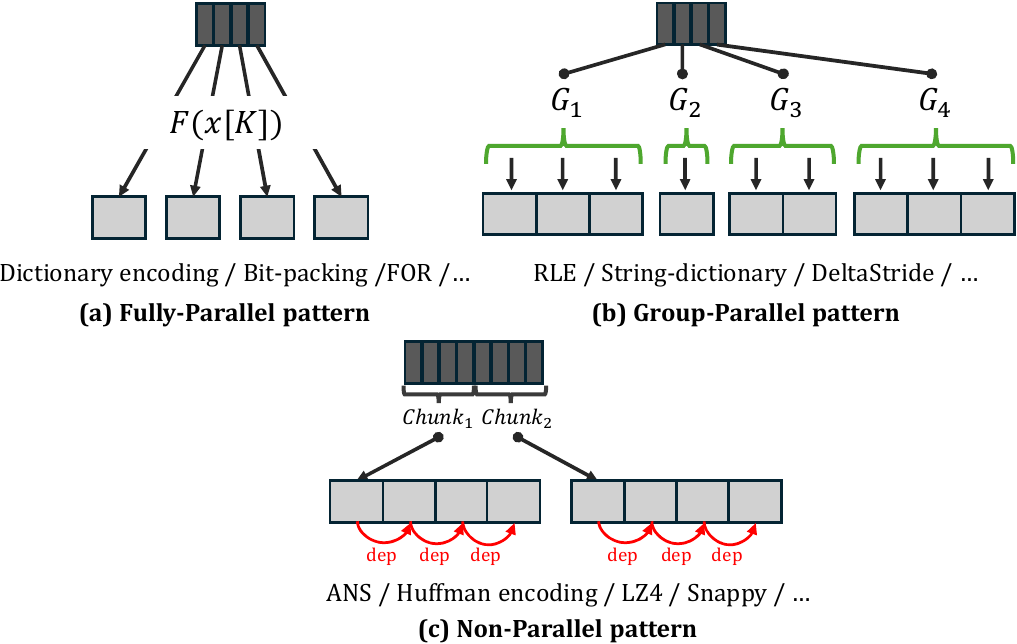}\\
	\caption{Logical dependencies for \proposed parallel patterns. 
    }
	\label{fig:design_patt}
\end{figure}

\subsection{Pattern Layer}
We identify three major parallel schemas based on our observations on commodity (de)compression algorithms.
To ensure comprehensive coverage of possible parallelism options, we classify the parallelizable schema into three distinct patterns: \emph{\pattA}, \emph{\pattB}, and \emph{\pattC}.
Given an input and output array, each pattern covers \textbf{N to 1}, \textbf{1 to N}, and \textbf{N to M} independent compute blocks in (de)compression algorithms, respectively.

\newcolumntype{L}{>{\arraybackslash}m{0.24\textwidth}}
\newcolumntype{P}[1]{>{\arraybackslash}p{#1}}
\newcolumntype{M}[1]{>{\arraybackslash}m{#1}}
\newcolumntype{Q}{>{\arraybackslash}m{0.31\textwidth}}

\begin{figure}[t!] \centering
\includegraphics[width=0.48\textwidth]{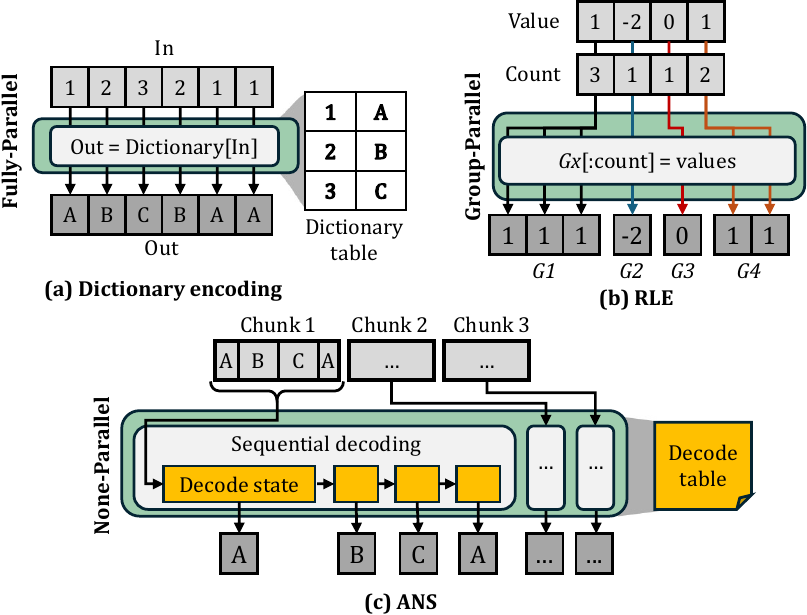}\\
	\caption{Data mapping example for each pattern.}
\label{fig:ops_example}
\end{figure}

\textbf{\pattA Pattern} is tailored for (de)compression operators where each input independently maps to output values with no cross-element dependencies, as shown in \fig{fig:design_patt}(a). Although the figure only shows a simple 1:1 scenario, a fixed scalar number of input arrays can exist to support the N to 1 computation block.
Any mapping function can be used to define the computation between input and output values, and arbitrary mappings between input and output indices are also supported.
Decompression techniques for algorithms such as dictionary encoding, bit-packing, float2int (G-ALP~\cite{galp}) and Fastlanes~\cite{afroozeh2024accelerating} algorithms, can be implemented using the \pattA kernel.
In \fig{fig:ops_example}(a), we present an example of dictionary-encoded data being decompressed using the \pattA kernel. The kernel uses a dictionary look-up table as a parallel mapping function, where the \texttt{Dictionary} is provided as metadata. This \pattA kernel performs parallel dictionary look-ups for each element selected by \texttt{In} and writes the corresponding value to \texttt{Out}.

\textbf{\pattB Pattern} focuses on customizing compression operators for spatially non-uniform distributions, where the entire task is segmented into multiple variable-sized groups $G_1,G_2,.., G_n$ containing independent subtasks within each group. Different sizes of $G_x$ result in an imbalance of execution. For example, \fig{fig:design_patt}(b) shows four groups, each containing a different number of elements internally, while the arrows within each group indicate a parallel relationship of 1 to N.
Typically, this pattern applies to all RLE-based encoding families including delta encoding, String-dictionary and {GSST~\cite{gsst}}.
We demonstrate how RLE decoding can be mapped to the \pattB pattern in \fig{fig:ops_example}(b). Compressed RLE data consists of a \texttt{count} array and a \texttt{value} array, where the \textit{count} array specifies the number of occurrences for each group, and each element in the \texttt{value} array corresponds to an input symbol. In this use case, a direct copy function is used as the mapping function when writing to the output, so that the \pattB kernel replicates each element in the \texttt{value} array in parallel, repeating it \texttt{count} times per group, and writes the results to the appropriate group indices.

\textbf{\pattC Pattern} is reserved for all other compression methods where parallelization is not feasible with \pattA and \pattB, such as inherently serial algorithms like Snappy and Huffman coding. We leverage the file chunking feature of the compression algorithm, dividing the target file into smaller chunks that are compressed and decompressed independently for parallel processing.
This approach enables different processors to handle independent file chunks concurrently, as illustrated in \fig{fig:design_patt}(c), allowing a single GPU processor to process serialized code as if it were implemented for a CPU.
Similar to many other serial algorithms, we take ANS decompression as an example. Each intermediate \texttt{decode state} in ANS depends on its predecessor to construct the corresponding element and recursively contributes its state to its successor. As shown in \fig{fig:ops_example}(c), this workflow enforces a strict, sequential progression of the \texttt{decode state} throughout the process within each chunk. However, opportunities for parallelism arise by grouping their intermediate \texttt{decode states} from different chunks and dispatching them in a SIMT (Single Instruction, Multiple Threads) manner, which is further explained in \sect{sect:implementation}.

\begin{figure}[t!] \centering
\includegraphics[width=0.48\textwidth]{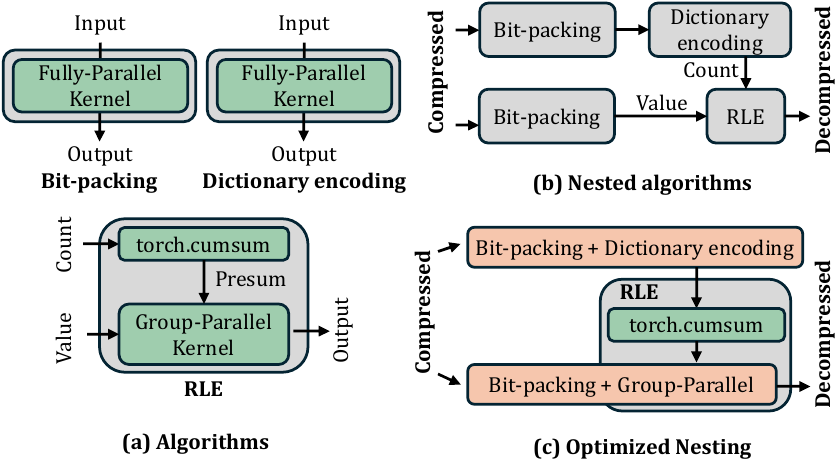}
\caption{Execution of Algorithm \& Nesting layer in PyTorch.}
\label{fig:framework_example}
\end{figure}

\subsection{Algorithm Layer \& Nesting Layer}
\label{sect:alg_nest}
The Algorithm Layer incorporates kernels compiled from the Pattern Layer to construct primitive (de)compression algorithms.
The pool of primitive (de)compression algorithms in \proposed includes delta encoding, dictionary encoding, String-dictionary, ANS, bit-packing, and Float2Int, while additional algorithms can be added when necessary, as demonstrated in \sect{sect:experiment}.
To further simplify the implementation, the framework also supports direct invocation of prebuilt operators from PyTorch’s out-of-the-box operations, such as sort, unique, and cumulative sum, as auxiliary operations.

\fig{fig:framework_example}(a) illustrates the decompression function design of bit-packing, dictionary encoding, and RLE. 
Decompression of bit-packing and dictionary encoding requires a single \pattA kernel, whereas RLE requires two-step kernels for decompression. The variable \texttt{count} is first pre-processed with PyTorch’s cumulative sum to calculate the base output index which is stored in a variable named \texttt{presum}. Next, the \pattB kernel is applied for parallel expansion of the compressed data. In the Nesting Layer, users can implement nested compression algorithms by combining a pool of primitive algorithms from the Algorithm Layer. These nested algorithms are regarded as the eventual (de)compression for target data columns. In \fig{fig:framework_example}(b), we show an example of nested decompression using sub-algorithms designed from \fig{fig:framework_example}(a),
while \fig{fig:framework_example}(c) further shows optimization by fusing \pattA patterns to nearby patterns. Consecutive \pattA kernels from the bit-packing and the dictionary encoding are fused to calculate the \texttt{count} tensor. 
Similarly, the bit-packing that generates the \texttt{value} tensor is also fused with the \pattB kernel inside the RLE function. Revisiting the Pattern Layer for kernel fusion enables further optimization by eliminating redundant memory movement.

\subsection{Pipelining Layer}

\begin{figure}[t!] \centering
\includegraphics[width=0.48\textwidth]{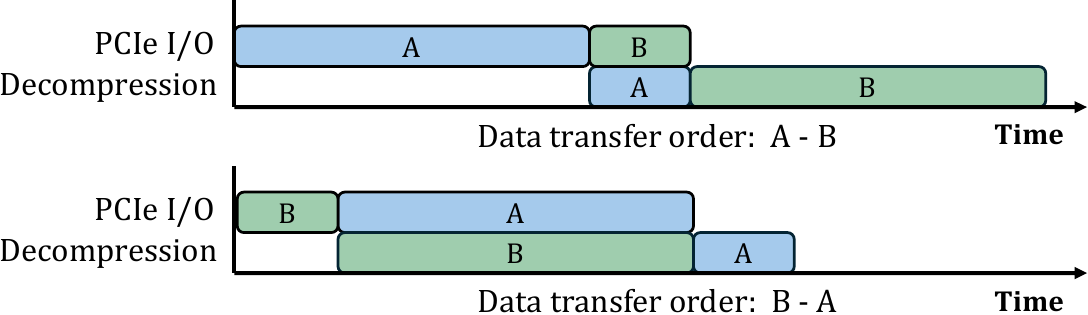}\\
	\caption{
    Transfer latency under different pipelining orders.
    }
    \vspace{-1.0em}
	\label{fig:pipeline}
\end{figure}

GPU-based query processing necessitates transferring multiple blocks of data from CPU$\rightarrow$GPU. To address the latency introduced by these data transfers and subsequent GPU-side decompression, \proposed introduces a dedicated Pipelining Layer. This layer aims to minimize end-to-end query latency by orchestrating the pipelined execution of CPU$\rightarrow$GPU data transfers alongside on-GPU decompression tasks, seamlessly overlapping these operations across several data blocks. Our empirical investigations reveal that the sequence in which data blocks are pipelined exerts a considerable impact on the achieved performance improvements. As illustrated in Figure~\ref{fig:pipeline}, consider two representative data blocks, A and B: data A is characterized by high data transfer overhead but rapid decompression, whereas data B exhibits the converse pattern, with low transfer cost but prolonged decompression time. By systematically comparing alternative data block transfer orders, we observe that initiating the pipeline with data B followed by data A (i.e., B$\rightarrow$A) achieves lower overall latency. This performance gain is due to the maximum concurrency between the data movement and decompression phases.

To generalize this optimization for arbitrary queries, it becomes critical to identify the optimal scheduling order of data transfers for minimum total latency. To this end, we design a specialized pipelining scheduler leveraging Johnson's algorithm~\cite{johnson1954optimal}. This scheduler efficiently determines the optimal data scheduling order with only $O(n)$ computational complexity.

\section{Device Geometry Scheduling}
\label{sect:implementation}
Enhancing the efficiency of the decompression kernel constitutes another crucial design aspect of \proposed. Given the diversity of GPU architectures in terms of on-chip resources such as compute units, memory transaction sizes, wavefront sizes, and cache capacities, compilation plays a key role in enabling fine-grained mapping of computational tasks to these heterogeneous native resources.
 We introduce a three-dimensional configuration vector, denoted as <\textit{L}, \textit{S}, \textit{C}>. Different tuple specifies a kernel instantiation that performs identical computation while leveraging distinct granularity of native hardware resources. This configurable format standardizes kernel scheduling by enabling precise adaptation of vector values to the architectural features of a specific GPU, thereby maximizing efficiency of both on-chip and off-chip resource utilization.

\begin{figure}[t!] \centering
\includegraphics[width=0.48\textwidth]{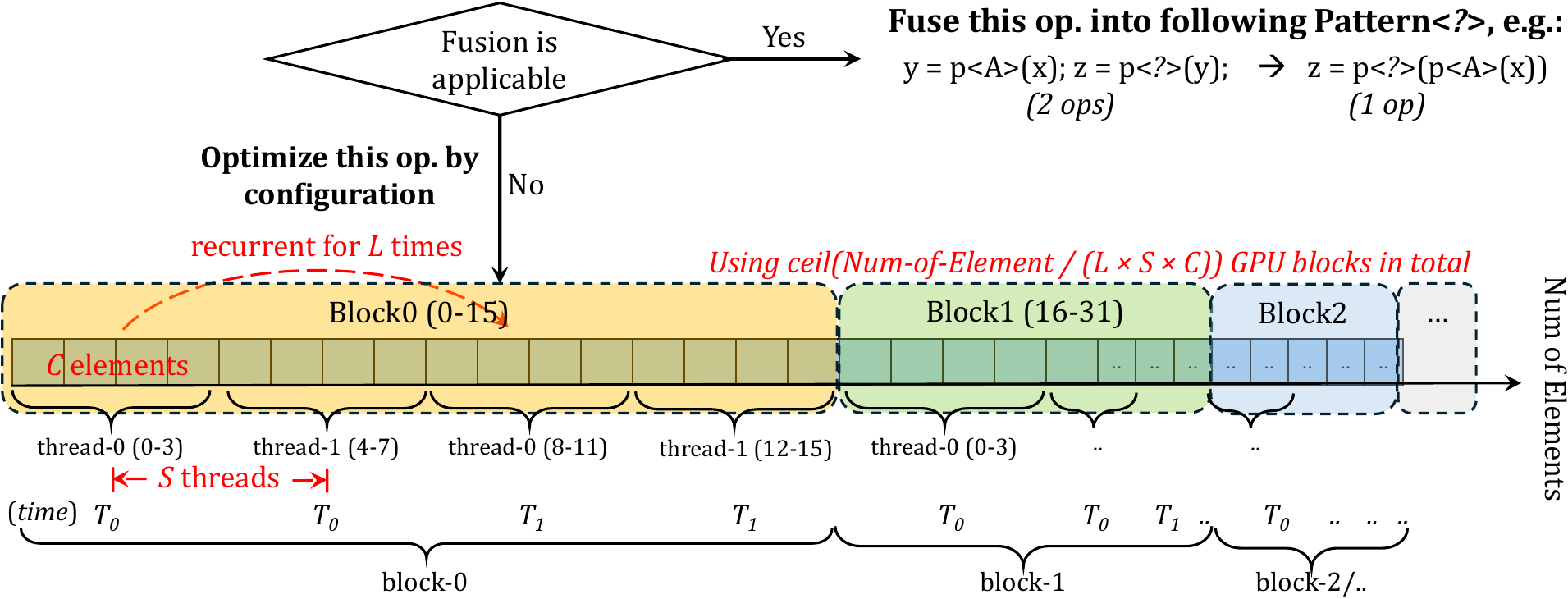}\\
	\caption{\pattA scheduling with <\textit{L},\textit{S},\textit{C}>=<2,2,4>.
    }
\label{fig:compile_pattA}
\end{figure}

\textbf{Scheduling \pattA with Fusion.} 
Compared with other computational patterns, the lack of cross-element dependencies endows the \pattA pattern with a unique capability for flexible kernel inlining and fusion. This property facilitates the integration of \pattA with a variety of other operations, such as seamlessly fusing type casting or dictionary lookups into RLE decoding, thereby improving overall computation efficiency and reducing kernel launch overhead. In scenarios where such fusion is infeasible, a separate GPU kernel must be employed during decompression tasks. To exemplify the kernel scheduling strategy for this pattern, \fig{fig:compile_pattA} presents a representative configuration. In this instance, the variable $C$ denotes the number of contiguous data elements jointly processed by a single GPU thread in one memory transaction, which is typically determined by the size of the data type under consideration. The parameter $S$ specifies the number of GPU threads allocated per block, dictating the degree of intra-block parallelism. $L$ indicates the iteration count of the main processing loop, such that each thread sequentially handles successive regions of data, offset by $S \times C$ elements, for $L$ iterations. Collectively, the product $L \times S \times C$ effectively defines the ``tile size'' that represents the total workload assigned to each GPU block. This flexible configuration not only matches the hardware execution model but also allows efficient scaling and adaptation across diverse GPU architectures.

\textbf{Scheduling \pattB for Load Balance.}
Efficiently processing non-uniform data on GPUs is challenging. Choosing RLE decompression as a typical example in this pattern, a straightforward strategy is to evenly distribute different RLE entries among all compute units for parallel execution. However, this approach often suffers from substantial load imbalance, since the workload of each subtask (i.e., the number of elements per RLE entry) can vary dramatically. As a result, some processors are assigned disproportionately heavy subtasks, while others complete early and remain idle, under-utilizing available resources.

\begin{figure}[t!] \centering
\includegraphics[width=0.48\textwidth]{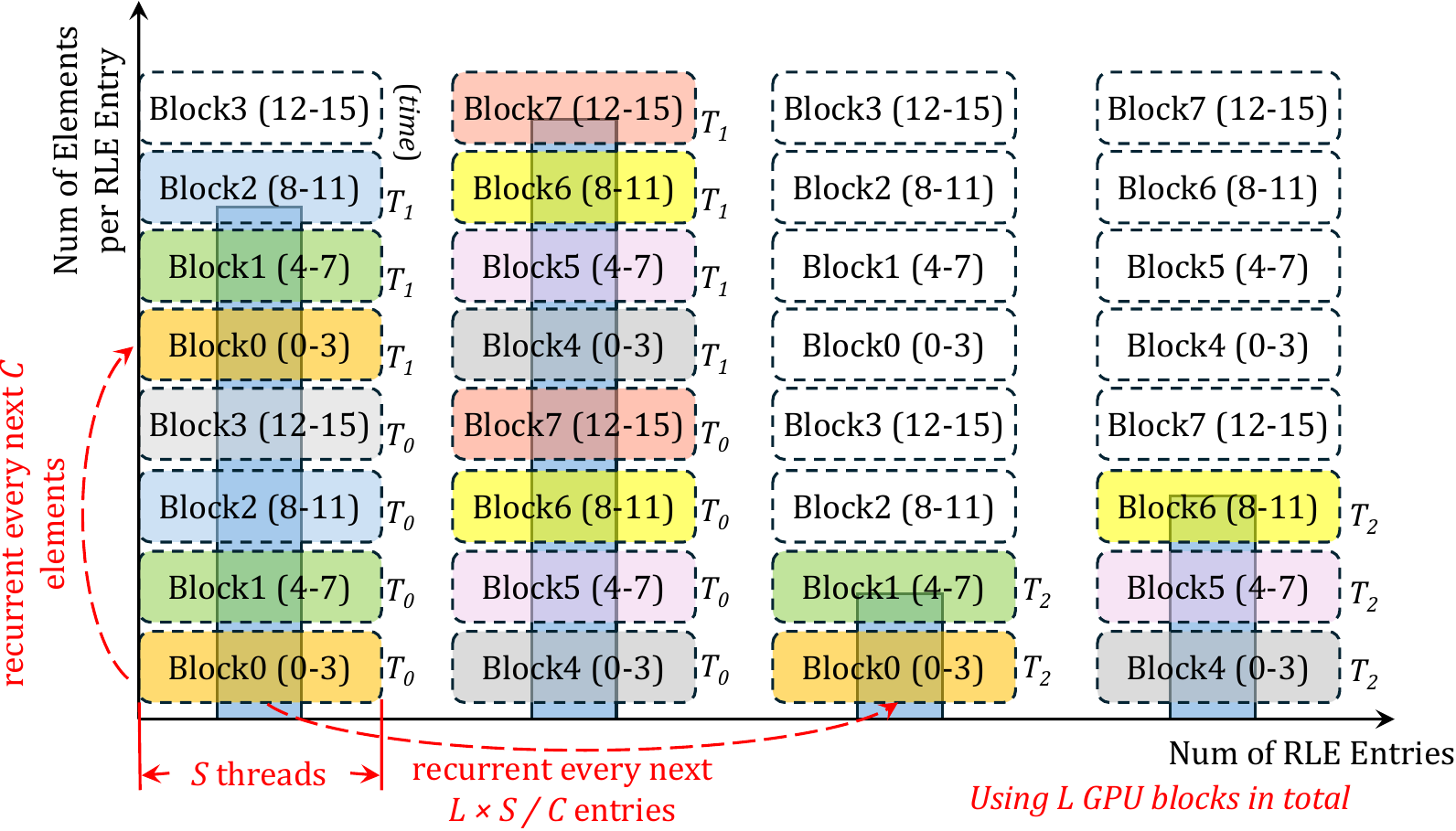}\\
	\caption{\pattB scheduling with <\textit{L},\textit{S},\textit{C}>=<8,4,16>.}
\label{fig:compile_pattB}
\end{figure}

\proposed tackles this divergence by combining \emph{multiple GPU blocks to co-process within a single group} and \emph{a single GPU block to process across multiple groups}. As illustrated in \fig{fig:compile_pattB}, the computation is organized for a four-group data arranged from left to right, each represented by a vertical bar whose height denotes the independent items within the group that can be processed in parallel. In this configuration, the entire data plane is processed by scheduling 8 GPU blocks (denoted as $L$), each containing 4 threads (denoted as $S$). For each group, 16 threads in total (denoted as $C$), evenly contributed by 4 GPU blocks (i.e., $C / S$), collaborate to process all the subtasks along the vertical axis. Along the horizontal dimension, each GPU block iteratively advances to process subsequent groups, skipping over a stride of 2 (i.e., $L \times S / C$) until all groups are exhausted. Notably, $S$ can exceed $C$, implying that surplus threads from a single GPU block may concurrently process tasks spanning across adjacent $S / C$ groups in a “lockstep” fashion, as opposed to strictly sequential execution from one group to the next. This schedule unblocks imbalanced data to scale along whatever vertical or horizontal dimension, which can also be achieved just by a one-time data scan.

\begin{figure}[t!] \centering
\includegraphics[width=0.48\textwidth]{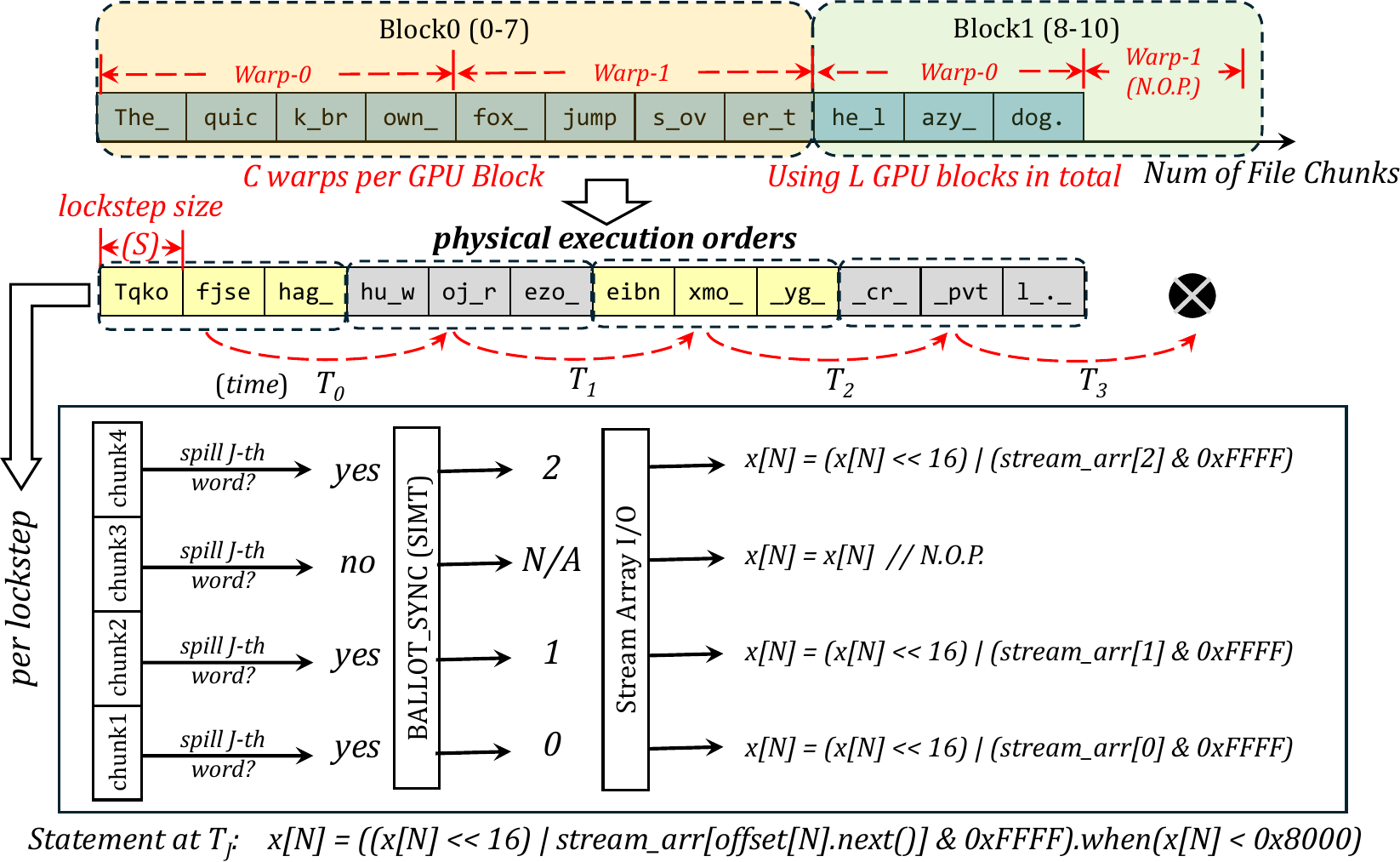}\\
	\caption{\pattC scheduling with <\textit{L},\textit{S},\textit{C}>=<2,4,2>.}
    \vspace{-1.0em}
\label{fig:compile_pattC}
\end{figure}

\textbf{Scheduling \pattC towards SIMT.} Given that the maximum achievable parallelism in this pattern is inherently bounded by the total number of data chunks, \proposed issues $L$ GPU blocks and $S \times C$ threads per block to deal with up to $L \times S \times C$ chunks of data, with each chunk managed by one thread for sequential execution. $L$ is calculated based on the number of chunks, while $S$ is fixed to the hardware warp size (e.g., 32 for all NVIDIA GPUs and 64 for most AMD GPUs). The parameter $C$ is architecture-dependent and enables each GPU block to handle more than $S$ chunks, depending on available register file size per physical compute unit. As illustrated in \fig{fig:compile_pattC}, which presents the physical execution order, a stack of sequential statements is repeatedly executed at different time steps without parallelism. Within each iteration, however, the processing of items from independent chunks are grouped into a SIMT unit, enabling lockstep execution. This approach enhances the consistency of I/O and cache accesses across chunks, thereby mitigating divergence and improving overall memory access efficiency.

\section{Experiment}
\label{sect:experiment}

\subsection{Experiment Setup}

We assess the compression ratio, decompression throughput, and end-to-end query execution latency of \proposed using the TPC-H benchmark. By default, all experiments use a TPC-H scale factor of 100 and are conducted on an NVIDIA A100 GPU~\cite{a100} with 80GB HBM2, connected to an AMD EPYC 7V12 64-core processor via PCIe Gen4 (32 GB/s theoretical bandwidth). To evaluate the device utilization of our approach across heterogeneous GPU architectures, we further conduct experiments on an NVIDIA H100 GPU~\cite{h100} with 80GB HBM2e, an AMD MI300X GPU~\cite{mi300x} with 192GB HBM3, and an AMD MI50 GPU~\cite{mi50} with 32GB HBM2. All measurements are performed with data buffers pinned to CPU memory, and we restrict execution to a single CPU node to minimize NUMA effects. We report the compression ratio as the quotient of compressed size over original size, and decompression throughput as the uncompressed data size divided by the decompression time.

For GPU (de)compression baseline, we employ nvCOMP v4.0.1.0 \cite{nvcomp}, which is widely recognized as the state-of-the-art compression library for NVIDIA GPUs. It provides a comprehensive suite of GPU-optimized algorithms, including standard methods such as LZ4, Snappy, Deflate, Zstandard (ZSTD), Asymmetric Numeral Systems (ANS), and Huffman encoding, as well as specialized techniques tailored for structured data (Cascaded), scientific data (Bitcomp), and GPU-optimized Deflate (GDeflate).
Among these, the Cascaded framework in nvCOMP supports configurable nesting of predefined run-length encoding (RLE), delta encoding, and bit-packing.

For the end-to-end TPC-H query execution experiment, we combine ``\proposed + TQP~\cite{tqp}'' to form a complete CPU-GPU hybrid query engine fully executed in the Pytorch environment (Python 3.12 + Pytorch 2.6.0). We further compare it with state-of-the-art CPU-based DBMS, specifically, DuckDB v1.4.1~\cite{duckdb} and Microsoft SQL Server 2022 RTM~\cite{sqlserver}, both of which were run {with 64 physical threads} on the same AMD processor as the GPU baseline.

\subsection{{Microbenchmark}}
\label{sect:pattern_experiment}
We begin with an apples-to-apples comparison against nvCOMP on the same compression patterns. For each primitive pattern, we select bit-packing, RLE, and ANS as representative instances and compare them to the corresponding algorithms in the nvCOMP library across a range of data distributions. {Given the closed-source nature of nvCOMP's implementations, we calibrate algorithm settings of \proposed and nvCOMP in terms of same compression ratio, thereby enabling a fair evaluation of performance metrics.}

\begin{figure}[t!] \centering
\includegraphics[width=0.48\textwidth]{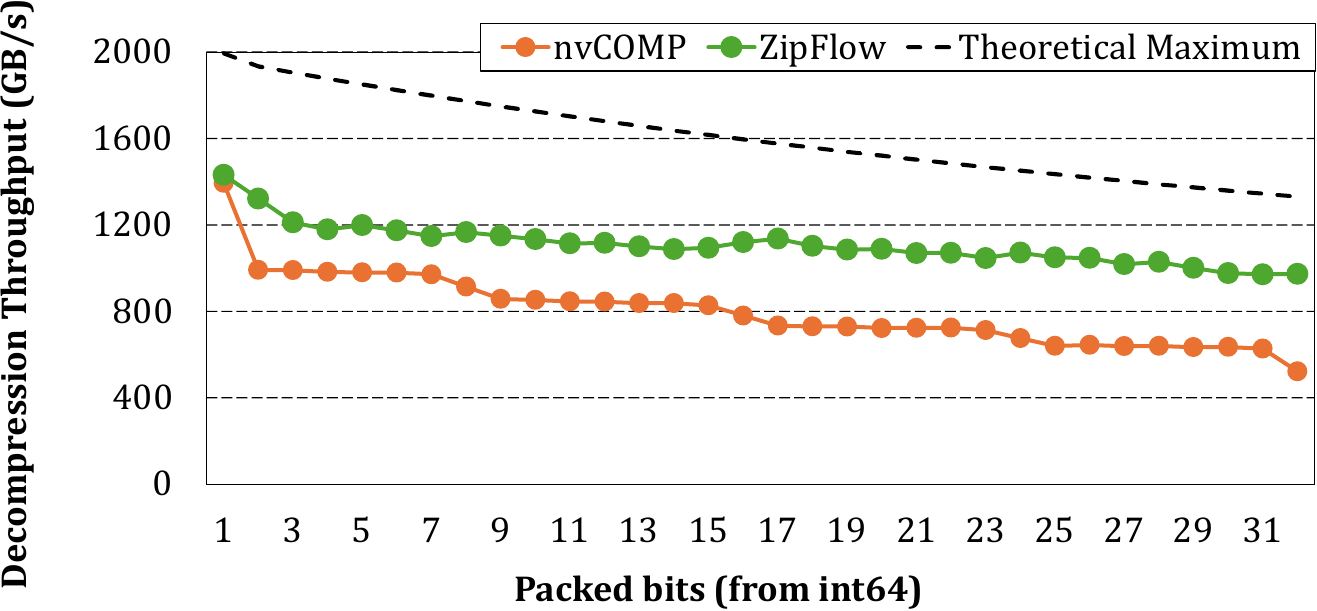}\\
	\caption{Throughput of bit-packing (\pattA) under varying distributions of bit-width to unpack.}
\label{fig:bitcompVSnvcomp}
\end{figure}

\subsubsection{\textbf{\textit{\pattA Instance: Bit-packing}}}
{\fig{fig:bitcompVSnvcomp} compares the decompression performance of the bit-packing implementations in \proposed and nvCOMP, with the dashed line for theoretical maximum throughput calculated by Equation \ref{eq:1}:

\begin{equation}
\label{eq:1}
GpuMemBandwidth\times\frac{\text{(plain size)}}{\text{(compressed size)} + \text{(plain size)}}
\end{equation}

To ensure a fair evaluation, we generate 4 GB synthetic input with different granularities of bit-width by uniformly sampling int64 values, with calibrated compression ratio.
Taking advantage of \proposed's native kernel scheduling for A100, it consistently outperforms nvCOMP as the compressed bit-width increases, and these results indicate that even for identical compression algorithms, a commodity library may not always deliver optimal performance depending on the input data pattern. For instance, on the \texttt{L\_PARTKEY} column in TPC-H (compressed to a 25-bit width), our method achieves a $1.63\times$ improvement in decompression performance.}

\begin{figure}[t!] \centering
\includegraphics[width=0.48\textwidth]{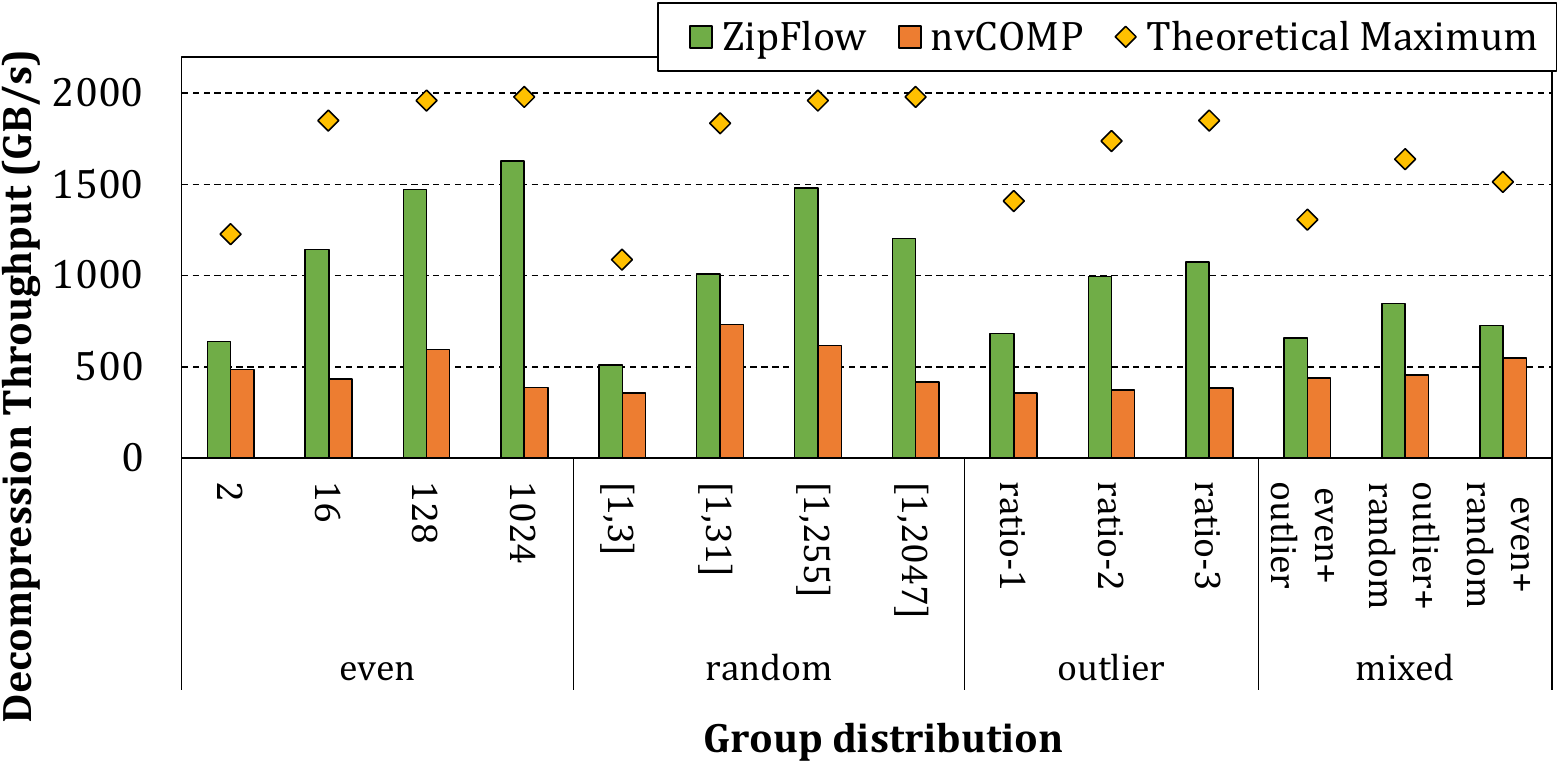}\\
	\caption{ Throughput of RLE (\pattB) under varying group size distributions: \textit{even}, \textit{random}, \textit{outlier}, \textit{mixed}.
    }
\label{fig:rleVSnvcomp}
\end{figure}

\subsubsection{\textbf{\textit{\pattB Instance: RLE}}}
We evaluate RLE decompression throughput across various group size distributions (i.e., the distribution of the \textit{count} array in RLE), considering four representative scenarios:
\textit{even}, \textit{random}, \textit{outlier}, and \textit{mixed}.
In the \textit{even-X} distribution, all groups contains identical number of elements, specifically of size X. For example, \textit{even-2} yields an uncompressed sequence such as A-A-B-B-C-C. The \textit{random}[L, R] distribution assigns each group a size selected uniformly at random from the range L to R.
The \textit{outlier} distribution consists predominantly of groups of size 1, interleaved with a small number of significantly larger groups (e.g., of size 1024), thereby simulating highly skewed group sizes. The \textit{mixed} distribution is generated by concatenating any two of the aforementioned scenarios, enabling the evaluation of performance under hybrid group size patterns.

In \fig{fig:rleVSnvcomp}, we provide a comprehensive performance comparison of RLE decompression using \proposed and nvCOMP across 14 representative data distribution cases. The bars represent the measured throughput, while the points denote the theoretical peak achievable under memory bandwidth constraints. The results show that nvCOMP offers limited adaptability to data distribution variability, maintaining consistently suboptimal performance that remains detached from the theoretical upper bound. This limitation primarily arises from nvCOMP's fixed parallelization approach, which assigns one thread per output element irrespective of group size, resulting in memory read contention and inefficient utilization of available bandwidth~\cite{nvcompGTC}. In contrast, \proposed maintains performance much closer to the theoretical maximum across all tested distributions. This close alignment demonstrates that \proposed effectively leverages both data locality and GPU parallelism, leading to improved occupancy and superior bandwidth utilization.

\begin{figure}[t!] \centering
\includegraphics[width=0.48\textwidth]{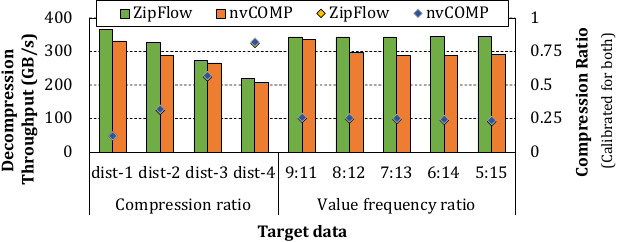}\\
	\caption{Throughput of ANS (\pattC) under varying compression ratio (left) \& frequency distributions (right). 
    }
\label{fig:ansVSnvcomp}
\end{figure}

\begin{figure}[t!] \centering
\includegraphics[width=0.48\textwidth]{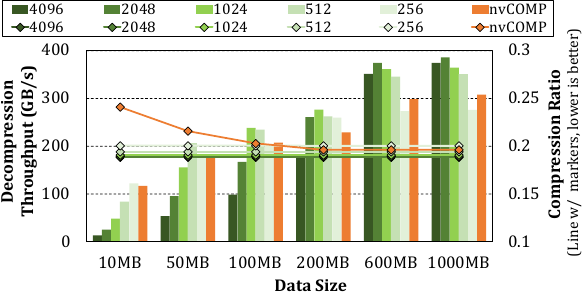}\\
	\caption{Throughput of ANS using varying chunk sizes (e.g., 4096 = 4KB per chunk, with nvCOMP under opaque setting).}
\label{fig:ansVSnvcomp_size}
\end{figure}

\subsubsection{\textbf{\textit{\pattC Instance: ANS}}}

For the \pattC pattern, we conduct the performance evaluation along two key dimensions: 1) varying the value frequency ratios to manipulate the entropy, as these significantly influence the compression efficacy of ANS (\fig{fig:ansVSnvcomp}); 2) analyzing the optimal chunk size across different total data volumes (\fig{fig:ansVSnvcomp_size}), as this parameter impacts both the compression ratio and decompression throughput.
Our evaluation datasets are constructed to capture the distributional properties of the \texttt{L\_RETURNFLAG} column from the TPC-H benchmark, where ANS compression is well-suited for encoding values containing distinct frequency distribution.

\fig{fig:ansVSnvcomp} presents the decompression throughput from two dimensions: four bars on the left depict decompression throughput across increasing compression ratios, while the remaining four bars on the right illustrate decompression throughput as the data frequency distribution shifts from balanced to imbalanced.
When the data distribution is made more skewed by increasing the frequency ratio of certain values, nvCOMP throughput drops noticeably.
In contrast, \proposed consistently maintains stable throughput, which has been observed in \pattA and \pattB, with its performance mostly governed by the compression ratio and shows minimal sensitivity to frequency variation or data skew.

\fig{fig:ansVSnvcomp_size} depicts a trade-off between employing larger chunk sizes, which yield greater compression ratios, and utilizing smaller chunk sizes that facilitate more parallelism and higher decompression throughput.
Given that the chunk size used by nvCOMP is not explicitly documented, we systematically sweep over a range of chunk sizes in the \proposed implementation to facilitate a calibrated comparison. Our findings reveal that, for smaller inputs, decreasing the chunk size leads to higher throughput by leveraging finer-grained parallelism. Conversely, for larger inputs, employing larger chunk sizes yields better scalability and improved throughput. Across all evaluated input sizes, our approach consistently identifies an optimal chunk size configuration that surpasses nvCOMP in terms of decompression throughput, while simultaneously achieving equal or better compression ratios.

In summary, the performance comparison among the three pattens (Fully-Parallel, Group-Parallel, and Non-Parallel) demonstrates the superior performance of \proposed. This improvement primarily arises from its fine-grained GPU scheduling adaptability. Such adaptability better aligns with both the characteristics of diverse data distributions and the architectural strengths of modern GPUs.

\newcolumntype{P}[1]{>{\centering\arraybackslash}m{#1}}
\begin{table}[t]
	\caption{Comparison between \proposed and public methods (nvCOMP~\cite{nvcomp}, Parquet~\cite{parquet}, BTRBLOCKS~\cite{btrblocks}), with the abbreviation of F.P., G.P., and N.P representing different pattern families: \pattA, \pattB and \pattC.}
    \scriptsize
    \centering
    \begin{tabular}{|P{0.005\textwidth} |P{0.1\textwidth}|P{0.05\textwidth}|P{0.05\textwidth}|P{0.08\textwidth}|P{0.05\textwidth}|}
    \hline
    \multicolumn{2}{|c|}{\textbf{Support}} & \textbf{nvCOMP} & \textbf{Parquet} & \textbf{BTRBLOCKS} & \textbf{\proposed}\\ \hline
    \multicolumn{2}{|c|}{Nesting} & Predefined & Predefined  & Extensible & Extensible\\ \hline
    \multicolumn{2}{|c|}{GPU support}  & \textcolor{green}{\cmark} & \textcolor{red}{\xmark} & \textcolor{red}{\xmark} & \textcolor{green}{\cmark}\\  \hline

    \multirow{4}{*}{\rotatebox[origin=c]{90}{ F.P.}} & Bit-pack \& FOR  & \textcolor{green}{\cmark} & \textcolor{green}{\cmark} & \textcolor{green}{\cmark} & \textcolor{green}{\cmark} \\ 
    & Delta encoding  & \textcolor{green}{\cmark} & \textcolor{green}{\cmark} & \textcolor{green}{\cmark}& \textcolor{green}{\cmark} \\ 
    & Dictionary encoding & \textcolor{red}{\xmark} & \textcolor{green}{\cmark} & \textcolor{green}{\cmark} & \textcolor{green}{\cmark} \\ 
    & Float2Int   & \textcolor{red}{\xmark} & \textcolor{red}{\xmark} & \textcolor{green}{\cmark}& \textcolor{green}{\cmark}\\ 
    \hline
    \multirow{5}{*}{\rotatebox[origin=c]{90}{ G.P.}} & RLE  & \textcolor{green}{\cmark} & \textcolor{green}{\cmark} & \textcolor{green}{\cmark} & \textcolor{green}{\cmark} \\
    & RLE + Delta   & \textcolor{green}{\cmark} & \textcolor{red}{\xmark}& \textcolor{green}{\cmark}& \textcolor{green}{\cmark} \\
    & RLE + Dictionary & \textcolor{red}{\xmark} & \textcolor{green}{\cmark}& \textcolor{green}{\cmark}& \textcolor{green}{\cmark} \\ 
    & String-dictionary  & \textcolor{red}{\xmark} & \textcolor{red}{\xmark}  & \textcolor{green}{\cmark} & \textcolor{green}{\cmark}\\
    & DeltaStride  & \textcolor{red}{\xmark} & \textcolor{red}{\xmark} & \textcolor{red}{\xmark} & \textcolor{green}{\cmark} \\
    \hline
    \multirow{3}{*}{\rotatebox[origin=c]{90}{N.P.}} & LZ4  & \textcolor{green}{\cmark} & \textcolor{green}{\cmark} & \textcolor{red}{\xmark} & \textcolor{green}{\cmark} \\
    & ANS   & \textcolor{green}{\cmark} & \textcolor{red}{\xmark}& \textcolor{red}{\xmark}& \textcolor{green}{\cmark} \\ 
    & Huffman & \textcolor{green}{\cmark} & \textcolor{red}{\xmark} & \textcolor{red}{\xmark} & \textcolor{green}{\cmark}  \\
    \hline
    \multicolumn{2}{|c|}{{Algorithm extensibility}}  & \textcolor{red}{\xmark} & \textcolor{red}{\xmark} & \textcolor{red}{\xmark} & \textcolor{green}{\cmark}\\\hline 
    \end{tabular}
    \label{tab:flex_comp}
\end{table}

\begin{figure*}[t!] \centering
\includegraphics[width=1\textwidth]{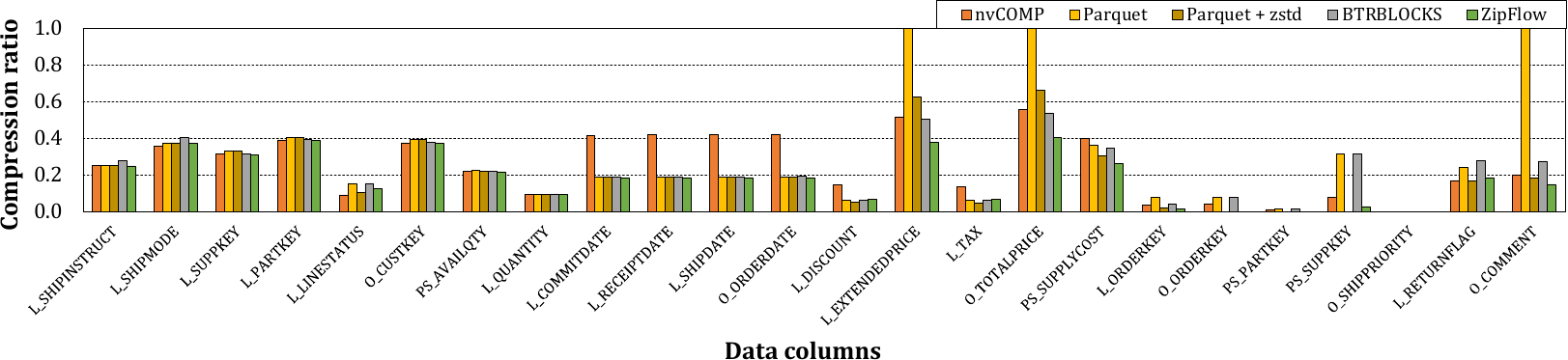}\\
    \vspace{-0.8em}
	\caption{Compression ratio comparison between \proposed and four different baselines. For Parquet and nvCOMP, we iterate through all possible algorithm options and report only the best ratio. For nvCOMP's cascaded, we iterate through 72 ($6\times6\times2$) different possible nesting levels of RLE, delta encoding, and bit-packing.}
    \vspace{-0.8em}
\label{fig:compratio_comparison}
\end{figure*}
\begin{figure*}[t!] \centering
\includegraphics[width=1\textwidth]{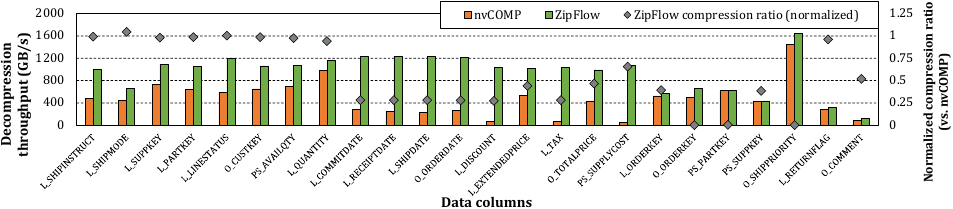}\\
	\captionsetup{justification=centering}
    \vspace{-0.8em}
    \caption{Decompression throughput comparison between nvCOMP and \proposed, where diamond markers represent \\ \proposed’s compression ratios normalized relative to nvCOMP.}
    \vspace{-0.8em}
\label{fig:compdecomp_all}
\end{figure*}

\subsection{{Compression Evaluation on \tpch Dataset}}
\label{sect:file_level_experiment}

For this evaluation, we pursue best possible performance for each data column, and design optimized compression algorithms by leveraging our framework’s nesting mechanisms and flexible kernel pool customization.
We focus on the three largest tables (\texttt{L}, \texttt{O}, and \texttt{PS}), which constitute the primary bottleneck for end-to-end query latency in TPC-H.
Our custom compression algorithms integrate nesting multiple techniques from different pattern families, such as bit-packing, RLE, delta encoding, dictionary encoding, ANS, and Float2Int. We also employ two specialized compression variants to handle particular data patterns. First, DeltaStride (a RLE variant) compresses monotonically increasing integer sequences by representing them as \textit{(start, stride, count)} triples.
Second, we generate a custom dictionary for the String‑dictionary compressor, the details of which are presented in \sect{sect:compratio}.
The nested compression algorithms we use are given in \tab{tab:applied_algs}.

\subsubsection{{\textbf{\textit{Compression Ratio}}}}
\label{sect:compratio}

In this section, we evaluate the compression ratios achieved by \proposed in comparison to one GPU-based baseline, nvCOMP~\cite{nvcomp}, and three CPU-based baselines: Parquet~\cite{parquet}, Parquet + zstd, and BTRBLOCKS~\cite{btrblocks},
irrespective of their compatibility for GPU.
\tab{tab:flex_comp} summarizes the set of compression algorithms supported by each framework, while \fig{fig:compratio_comparison} presents their compression ratios. For both nvCOMP and Parquet, we explore their best compression ratio attained across all supported algorithms listed in \tab{tab:flex_comp}. We further provide a detailed analysis of the results for representative column clusters below.

As shown in \fig{fig:compratio_comparison}, columns containing date values (i.e., four columns plotted in between \texttt{L\_COMMITDATE} and \texttt{O\_ORDERDATE}) benefit substantially from dictionary encoding. In contrast, nvCOMP does not support this data representation and therefore yields consistently lower compression ratios relative to all alternative solutions. For numerical columns (i.e., five columns plotted from \texttt{L\_DISCOUNT} to \texttt{PS\_SUPPLYCOST}), our framework employs a custom Float2Int compression technique from the \pattA family, which outperforms other methods in terms of compression efficiency. 
For columns storing primary keys (e.g., \texttt{O\_ORDERKEY}), we adopt a hybrid approach that combines lightweight methods: the nesting of DeltaStride and the bit-packing algorithm respectively from \pattB and \pattA families. This strategy is well-suited for encoding ``nearly'' monotonically increasing integer sequences, resulting in substantially improved compression ratios.
For string-valued columns like \texttt{O\_COMMENT}, we introduce a tailored dictionary construction from the \pattB family by tokenizing on spaces and periods. This yields a compact String-dictionary encompassing $1,878$ unique words, where indices can be further bit-packed into 12 bits. When decompressing, each unique word serve as a group in \pattB and expands according to the lookup dictionary.
The String-dictionary approach avoids the expensive decompression overheads of LZ77  (\pattC) algorithms.
We further cascade our custom String-dictionary encoding with bit-packing and ANS. This approach consistently achieves higher compression ratios than both nvCOMP (zstd) and BTRBLOCKS (FSST\cite{fsst}), as reflected in our experimental results.

\newcolumntype{L}{>{\arraybackslash}m{0.31\textwidth}}
\newcolumntype{P}[1]{>{\centering\arraybackslash}m{#1}}
\begin{table}[t]
	\caption{Custom compression algorithm for each file. “|” divides each nested algorithm. Algorithms with two or more outputs are enclosed with “[,]”. 
    }
    \scriptsize
    \centering
    \begin{tabular}{|p{0.12\textwidth} | L |}
    \hline
    \textbf{Data Columns }& \textbf{Nested algorithms} \\
    \hline
    \texttt{L\_SHIPINSTRUCT} & Bit-packing  \\ \hline
    \texttt{L\_SHIPMODE} & Bit-packing \\ \hline
    \texttt{L\_SUPPKEY} & Bit-packing \\ \hline
    \texttt{L\_PARTKEY} & Bit-packing \\ \hline
    \texttt{L\_LINESTATUS} & Bit-packing  \\ \hline
    \texttt{O\_CUSTKEY} & Bit-packing  \\ \hline
    \texttt{PS\_AVAILQTY} & Bit-packing \\ \hline
    \texttt{L\_QUANTITY} & Bit-packing \\ \hline
    \texttt{L\_COMMITDATE} & Dictionary encoding | Bit-packing  \\ \hline
    \texttt{L\_RECEIPTDATE} & Dictionary encoding | Bit-packing  \\ \hline
    \texttt{L\_SHIPDATE} &  Dictionary encoding | Bit-packing   \\ \hline
    \texttt{O\_ORDERDATE} &  Dictionary encoding | Bit-packing   \\ \hline
    \texttt{L\_DISCOUNT} &  Float2Int | Bit-packing   \\ \hline
    \texttt{L\_EXTENDEDPRICE} &  Float2Int | Bit-packing   \\ \hline
    \texttt{L\_TAX} &  Float2Int | Bit-packing  \\ \hline
    \texttt{O\_TOTALPRICE} &  Float2Int | Bit-packing   \\ \hline
    \texttt{PS\_SUPPLYCOST} &  Float2Int | Bit-packing   \\ \hline
    \texttt{L\_ORDERKEY} & RLE | [DeltaStride | [Delta encoding | RLE | [Bit-packing,Bit-packing], Bit-packing], Bit-packing]~]  \\ \hline
    \texttt{O\_ORDERKEY} & DeltaStride | [Delta encoding | RLE | [Bit-packing,Bit-packing], Bit-packing]  \\ \hline
    \texttt{PS\_PARTKEY} & RLE | [DeltaStride , RLE ] ]  \\ \hline
    \texttt{PS\_SUPPKEY} & Delta encoding | Dictionary encoding | Bit-packing | Dictionary encoding | Bit-packing \\ \hline
    \texttt{O\_SHIPPRIORITY} & RLE  \\ \hline
    \texttt{L\_RETURNFLAG} & ANS \\ \hline
    \texttt{O\_COMMENT} & String-dictionary | Bit-packing | ANS  \\ 
    \hline
    \end{tabular}

    \label{tab:applied_algs}
\end{table}

Overall, \proposed broadens the range of possible nesting combinations among a wider range of custom algorithm choices, which is a key factor in discovering better compression schemes across different column types.

\subsubsection{{\textbf{\textit{Decompression Throughput}}}}
\label{sect:decomp_throughput}

This section evaluates GPU decompression throughput and its influence on file-level data movement. In contrast to the compression experiments, nvCOMP serves as the sole state-of-the-art baseline for comparison, since
Parquet and BTRBLOCKS lack the capability to perform decompression after the compressed data are transferred to GPU.

From the nvCOMP library, we select the algorithm that delivers the highest end-to-end performance by minimizing combined I/O and decompression latency among the functions provided. \fig{fig:compdecomp_all} presents the decompression throughput of nvCOMP and \proposed across all data columns, and additionally illustrates the compression ratio advantage of \proposed over nvCOMP as indicated by the gray diamond markers.
The leftmost eight data points correspond to columns based on bit-packing. For these columns, the optimal algorithm from nvCOMP is the bit-packing implementation within the Cascaded function, which yields the same compression ratio as our method. 
For columns ranging from \texttt{L\_COMMITDATE} to \texttt{PS\_SUPPLYCOST}, nvCOMP exhibits notably poor performance. Since nvCOMP lacks support for the dictionary encoding and Float2Int algorithms, it resorts to various alternative techniques to optimize data movement, such as Bitcomp, Gdeflate, ANS, and Cascaded. 
However, these complex algorithms fail to provide advantages in terms of either compression ratio or decompression throughput, as shown in \fig{fig:compdecomp_all}.
In contrast, \proposed leverages a specialized yet lightweight \pattA solution, achieving on average a $21\%$ improvement in compression ratio while delivering substantially higher decompression throughput. In summary, \proposed achieves an average reduction in file-level data movement overhead by a factor of $2.07\times$ relative to nvCOMP.

\subsubsection{\textbf{\textit{Impact of Fusion on Decompression}}}

\begin{figure}[t!] \centering
\includegraphics[width=0.48\textwidth]{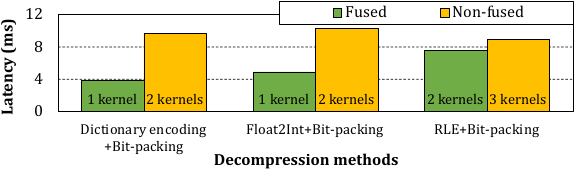}\\
	\caption{Kernel fusion on nested decompression kernels.}
    \vspace{-1.0em}
\label{fig:fused_ablation}
\end{figure}

We conduct an ablation study to evaluate the impact of kernel fusion on nested decompression. 
In \fig{fig:fused_ablation}, we present the latency of fused and non-fused kernels on three nested functions: Float2Int + Bit-packing, Dictionary encoding + Bit-packing, and RLE + Bit-packing. Each algorithm is applied to \texttt{L\_SHIPDATE}, \texttt{L\_EXTENDEDPRICE}, and \texttt{L\_ORDERKEY}, respectively, where all three files have the same size. The fused version of the kernel for Dictionary encoding + Bit-packing and Float2Int + Bit-packing executes a single kernel, while RLE + Bit-packing executes two kernels (i.e., PyTorch's cumulative sum and \proposed operator). The non-fused version does not fuse the bit-packing operator, resulting in two, two, and three separate kernel executions.

Empirically, kernel fusion yields a performance improvement of $2.45\times$ for Dictionary encoding + Bit-packing and $2.08\times$ for Float2Int + Bit-packing over their non-fused counterparts. This significant performance gap arises from the variances in off-chip memory access. In Equation~\ref{eq:2}, we explain the difference of required memory traffic of fused and non-fused kernels for above two cases. 
Memory traffic of single fused kernel consists of only the \textit{compressed size} and \textit{plain size} (i.e. uncompressed size), while non-fused kernel require extra memory round trip of  intermediate value. As dictionary encoding and Float2Int alone do not compress the data, the intermediate data cost is considered as \textit{plain size}.

\begin{equation}
\label{eq:2}
\frac{1\times\text{(compressed size)} + 3\times\text{(plain size)}}{1\times\text{(compressed size)} + 1\times\text{(plain size)}} > 2
\end{equation}

This result underscores the substantial penalty incurred by even a single additional memory round trip in non-fused pipelines. 
In contrast, we observe only a moderate speedup for RLE + Bit-packing, where kernel fusion increases performance by $1.18\times$ relative to the non-fused baseline. This is because bit-packing is only applied to the volume \texttt{count}, which has already been compressed by RLE and only constitutes $12.5\%$ of the original data. As a result, the fused kernel achieves a limited performance boost.

\begin{figure*}[t!] \centering
\includegraphics[width=1\textwidth]{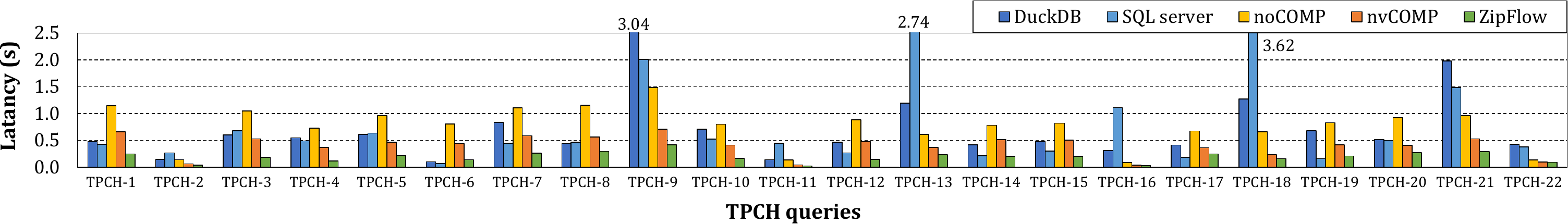}\\
	\captionsetup{justification=centering}
    \vspace{-0.8em}
    \caption{{TPC-H query latency comparison with different CPU and GPU baselines.}
    }
    \vspace{-0.8em}
\label{fig:query_level_all}
\end{figure*}

\begin{figure*}[t!] \centering
\includegraphics[width=1\textwidth]{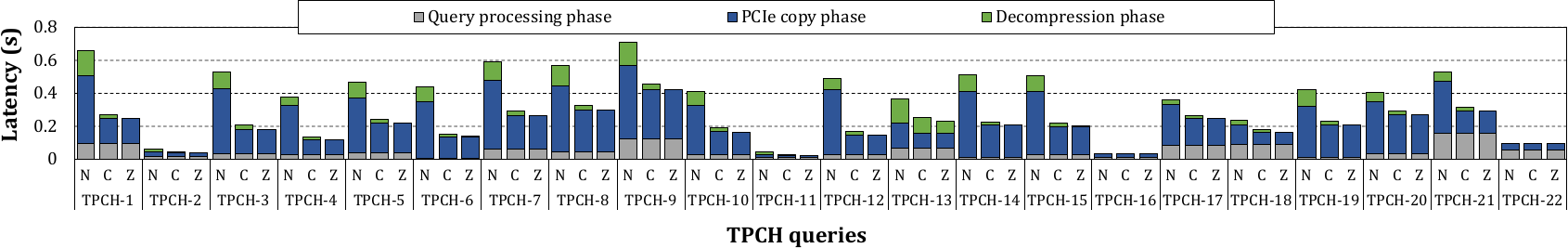}\\
    \vspace{-0.8em}
    \caption{Breakdown of TPC-H query latency by execution phases —— ``N'' for nvCOMP, ``C'' for \proposed with only (de)compression applied, and ``Z'' for full \proposed implementation including PCIe pipelining.}
    \vspace{-0.8em}
\label{fig:query_all_breakdown}
\end{figure*}

\begin{figure}[t!] \centering
\includegraphics[width=0.48\textwidth]{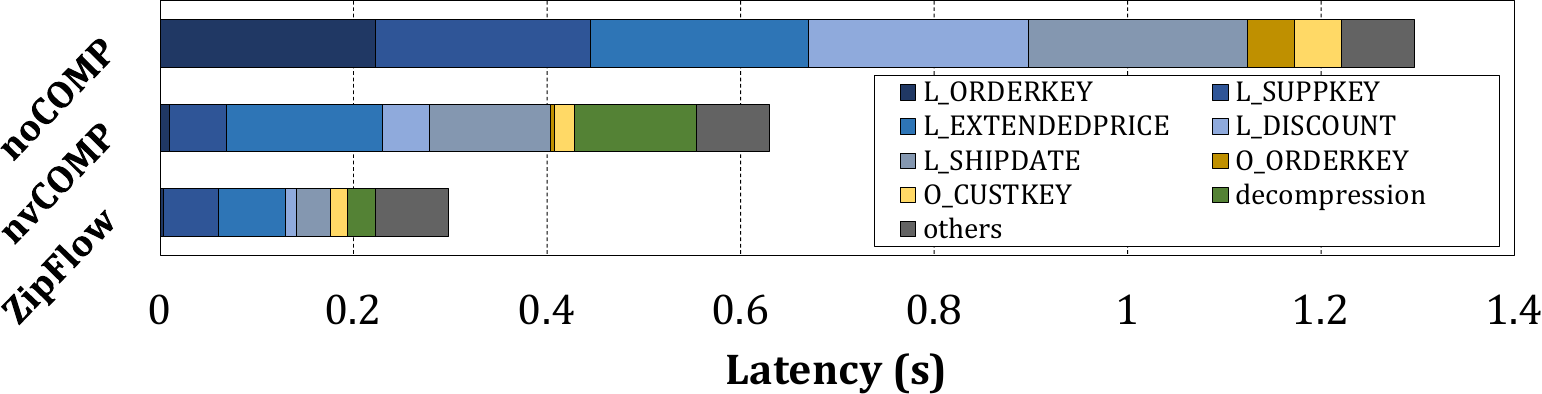}\\
	\caption{Breakdown of TPC-H query 7's execution latency by each input column
    (“others” refer to the cost spent on query processing kernels and uncompressed data transfer).}
    \vspace{-1.0em}
\label{fig:query7_breakdown}
\end{figure}

\subsection{End-to-end TPC-H Query Performance}
\label{sect:query_level_experiment}

We benchmark \proposed against several baselines in a full-pipeline, end-to-end query execution setting using the complete TPC-H workload (all 22 queries). For each query, we transfer only the columns required by that query to the GPU. Compression is applied exclusively to the columns listed in Table \ref{tab:applied_algs}, while other columns such as nation ID, are small enough that compression would not provide a measurable benefit. After the required data is prepared in GPU memory, we synthetically incorporate query execution latencies measured by TQP~\cite{tqp} on the same A100 GPU. To ensure a fair comparison of end-to-end costs and to equalize the cost of the query processing phase, we employ TQP as the query processing engine across all GPU-based baselines, while varying the compression configurations as follows: (1) an approach where no compression is applied (labeled ``noCOMP'') and (2) applying the nvCOMP library with the best-performing algorithms identified in \sect{sect:decomp_throughput}.
For the CPU-based baseline, we compare against state-of-the-art database systems, DuckDB and Microsoft SQL Server.

\subsubsection{{\textbf{\textit{Overall Query Latency}}}}
Figure \ref{fig:query_level_all} illustrates the per-query execution times for all configurations. The baseline noComp suffers the highest latency because PCIe data transfers dominate the runtime, accounting for an average of $91.3\%$ of the total execution time.
Consequently, both nvCOMP and \proposed dramatically cut the PCIe overhead down, yielding latency reductions of $1.91\times$ and $4.08\times$, respectively. Across the entire benchmark, \proposed consistently outperforms nvCOMP, delivering an average end-to-end speedup of $2.08\times$.

While GPU-based solutions demonstrate significant performance advantages in most cases, CPU baselines remain competitive for several queries:
DuckDB and SQL Server outperform the nvCOMP baseline respectively in 6 and 9 out of 22 queries. 
This occurs because the data movement overheads of nvCOMP outweigh the benefits provided by GPU acceleration. 
Based on a deeper data transfer optimization with \proposed, GPU-based query execution outperforms DuckDB in all queries except TPC-H 6, and outperforms SQL Server in all but three queries (TPC-H 6, 17, and 19). 
Queries such as TPC-H 6 are dominated by a 1-time scan operation and primarily constrained by memory and PCIe bandwidth. Since CPU memory bandwidth far exceeds PCIe bandwidth, these queries are more suitable for CPU execution and gain little benefit from transferring data to the GPU.
Overall, \proposed delivers average speedups of $3.14\times$ over DuckDB and $3.32\times$ over SQL Server.

\subsubsection{\textbf{\textit{Latency During GPU Decompression}}}
{\fig{fig:query_all_breakdown} decomposes the query‑execution latency of \proposed and nvCOMP into their constituent parts. To make the latency contributions of each part clearer, we plot the bar with label ``C'' that stands for \proposed without pipelining the copy and decompression phases}, which clearly shows the latency saved by overlapping those operations.
On average, our approach achieves a $3.26\times$ reduction in GPU kernel latency. Among the 22 queries, we observed that those including the \texttt{L\_TAX} and \texttt{L\_DISCOUNT} columns (TPCH 1, 3, 5–10, 14, 15, and 19) exhibit significant differences in GPU decompression latency between our solution and nvCOMP. This gap arises because the optimal algorithm for \texttt{L\_TAX} and \texttt{L\_DISCOUNT} in nvCOMP is Gdeflate, which has a relatively low decompression speed of an average of 66.6GB/s. In contrast, our proposed solution employs specialized Float2Int decompression, achieving an average speed of 1030GB/s, effectively reducing decompression overhead.

\subsubsection{\textbf{\textit{Latency During PCIe I/O}}}

Attributed to the reduced size of compressed data transferred, \proposed achieves a substantial reduction in PCIe I/O latency, resulting in $1.85\times$ improvement than that achieved by nvCOMP on average. 
Since PCIe transfer time is highly sensitive to compression ratio, the observed improvement is primarily driven by columns compressed using dictionary encoding and the Float2Int algorithm, for which our framework demonstrates a significant compression ratio advantage (shown in \fig{fig:compdecomp_all}).
To provide a closer examination, we select TPCH-7 as a representative example and present a further detailed latency breakdown of three different execution scenarios in \fig{fig:query7_breakdown}. Our observations reveal that, in the nvCOMP-based compression strategy for TPCH-7, \texttt{L\_EXTENDEDPRICE} and \texttt{L\_SHIPDATE} contribute most significantly to I/O latency.
In contrast, our proposed solution employs compression methods optimized for these columns, resulting in a more balanced latency distribution across all columns.
Interestingly, even though \texttt{L\_ORDERKEY} and \texttt{L\_SUPPKEY} have the same original size as the aforementioned columns, they exhibit a far smaller impact on end-to-end latency. This is due to their high compression ratio and decompression speed, resulting in only a small portion of the overall query latency. A similar pattern is observed in TPCH-13, where latency is dominated by the \texttt{O\_COMMENT} column, which constitutes $89.9\%$ of the total data transfer volume.
From the above observations, we conclude that as compressed data movement is an effective solution, the lack of compression support for a few data patterns becomes a noticeable outlier. Specifically, when other columns of similar size are efficiently compressed and decompressed, unsupported data patterns can result in disproportionate latency, making certain columns stand out as bottlenecks in end-to-end query performance.

\subsubsection{\textbf{\textit{Impact of PCIe/GPU Pipelining.}}}
We analyze the impact of the Pipelining layer by comparing \proposed with and without pipelining, labeled as "Z" and "C" in \fig{fig:query_all_breakdown}, respectively.
For all queries except TPCH-13, pipelining significantly reduces the exposed decompression kernel latency, resulting in an average of 10\% reduction in end-to-end query latency compared to non-pipelined execution. 
We observe that, in our current compression method design, all columns except RLE/Delta-relevant compression are dominated by PCIe I/O latency. This allows for sufficient overlap of most GPU decompression latency, but also indicates that PCIe I/O is still the primary bottleneck in the full query pipeline. This observation highlights a key optimization opportunity for designers: by further nesting complex compression methods to achieve higher compression ratios, PCIe I/O bottlenecks can be mitigated at the cost of decompression throughput. It is also important to note that the tradeoff between I/O and decompression latencies is largely determined by the bandwidth disparity between PCIe and GPU memory, which represents an interesting direction for future work.

\subsection{Evaluation on Heterogeneous GPUs}
\label{sect:kernel_level_experiment}

\begin{figure}[t!] \centering
\includegraphics[width=0.48\textwidth]{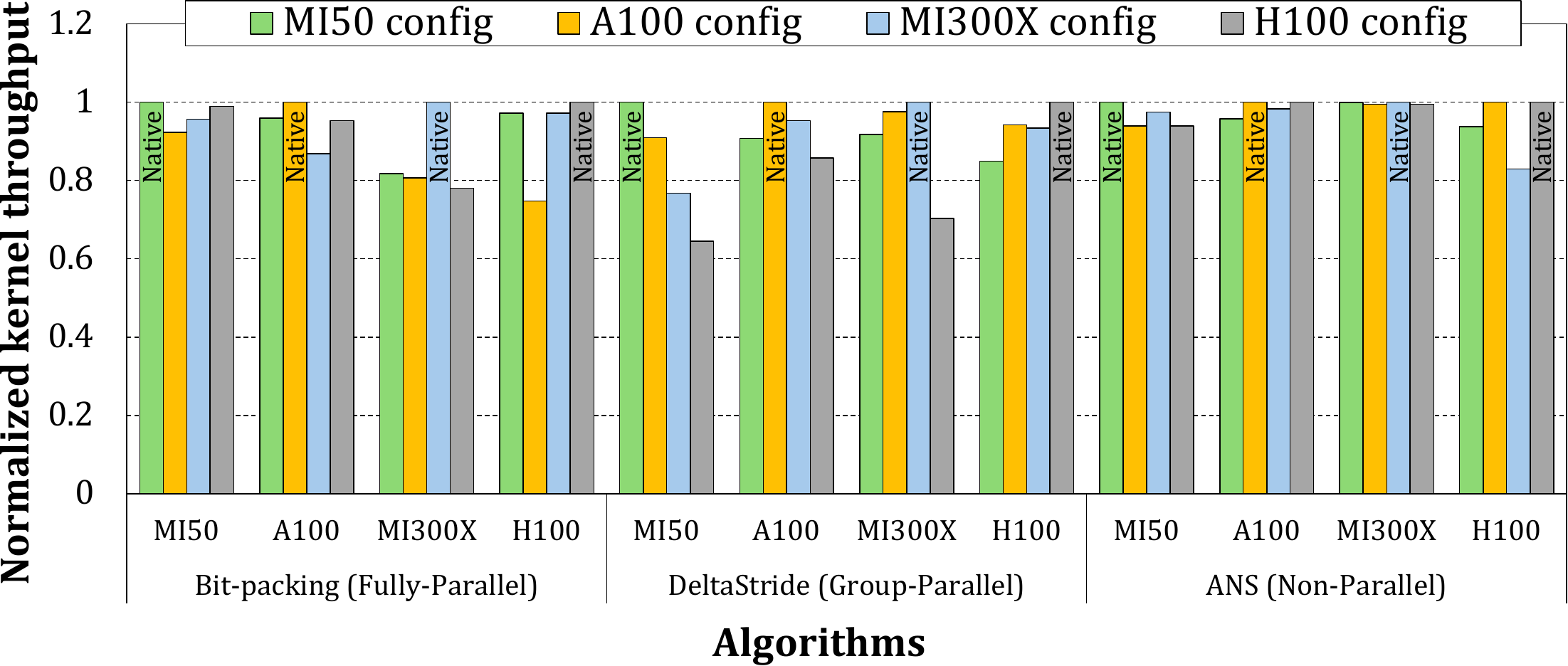}\\
	\caption{Kernel efficiencies under ``Native Config'' and ``Shared Config'' across 4 GPU types.  All values are normalized to the respective kernel efficiency in the ``Native Config''.}
    \vspace{-1.4em}
\label{fig:diff_hw_kernels}
\end{figure}

In this section, we analyze the impact of \proposed configurations across four different types of GPUs, along with the overhead incurred when identifying optimal candidates from the configuration space. Although GPUs differ in terms of core counts, cache capacity, and other architectural resources, \proposed enables a kernel designed for one GPU to be executed on another. However, employing a non-optimal configuration typically results in sub-optimal decompression performance. Given that the optimal kernel configuration may vary across hardware platforms, we define the configuration tuned for a specific GPU as its ``Native Config'', and refer to configurations originally designed for other GPUs as ``Shared Configs''. As illustrated in \fig{fig:diff_hw_kernels}, we first fine-tune the necessary operators on four representative GPUs (MI50~\cite{mi50}, A100~\cite{a100}, H100~\cite{h100}, and MI300x~\cite{mi300x}), obtaining four corresponding ``Native Configs''. We then test each of these configurations on the remaining GPUs (as ``Shared Config'') to assess their performance impact. Across all scenarios, we observe that a GPU employing a ``Shared Config'' is unable to fully realize its inherent performance potential, with degradations of up to 35\%.

In real-world data analytics scenarios, an efficient offline tuning approach is required to determine the optimal “Native Config” in advance, thereby preventing performance degradation associated with using a “Shared Config” at runtime.
We reduce offline configuration exploration costs through reinforcement learning within a carefully narrowed exploration space, where all values are restricted to powers of 2 which are presented in \tab{tab:exploration_cost}. As a baseline, “B.F Search'' (alias. Brute-force Search) means exploring all combinations of candidates that are valid in the space. In contrast, “R.L. Search'' (alias. Reinforcement learning Search) leverages the hidden monotonicity in the performance distribution to prune the search space, thereby further reducing the actual cost of candidate exploration down to within ten.

\renewcommand{\arraystretch}{1.2}
\newcolumntype{P}[1]{>{\centering\arraybackslash}p{#1}}
\begin{table}[t]
	\caption{{Pattern exploration cost on NVIDIA and AMD GPUs. We label the pattern types as F.P.(\pattA), G.P.(\pattB), and N.P.(\pattC).}}

    \scriptsize
    \centering
    \begin{tabular}{|P{0.01\textwidth}|P{0.08\textwidth}|P{0.07\textwidth}|P{0.12\textwidth}| P{0.08\textwidth}|}
    \hline
    \multicolumn{2}{|c|}{} & $range(L)$ & $range(S)$ & $range(C)$ \\ \hline 
    \multirow{3}{*}{\rotatebox[origin=c]{90}{ F.P.}} & Config space & $2^0 .. 2^4$  & $Hw.WarpSize .. 2^{10}$ &  $\lceil4/dtype.size\rceil$  \\ \cline{2-5}
    & B.F. Search & \multicolumn{3}{c|}{\(NVIDIA = 5 \times 6 \times 1\) ,\quad  \(AMD = 5 \times 7 \times 1\)} \\ \cline{2-5}
    & R.L. Search & \multicolumn{3}{c|}{\(NVIDIA \approx 3 + 4 + 0\)    , \quad    \(AMD \approx 3 + 4 + 0\)} \\ \hline
    \multirow{3}{*}{\rotatebox[origin=c]{90}{ G.P.}} & Config space & $Hw.numCUs$  & $Hw.WarpSize..2^{10}$ & $2^0 .. 2^{10}$ \\ \cline{2-5}
    & B.F. Search& \multicolumn{3}{c|}{\(NVIDIA = 1 \times 6 \times 11\) ,\quad \(AMD = 1 \times 7 \times 11\)} \\ \cline{2-5}
    & R.L. Search& \multicolumn{3}{c|}{\( NVIDIA \approx 0 + 4 + 5\) ,\quad  \(AMD \approx 0 + 4 + 5\)} \\  \hline
    \multirow{3}{*}{\rotatebox[origin=c]{90}{ N.P.}} & Config space & $Hw.numCUs$  & $Hw.WarpSize$ & $2^0 .. 2^{10}$ \\ \cline{2-5}
    & B.F. Search& \multicolumn{3}{c|}{ \(NVIDIA = 1 \times 1 \times 11\) ,\quad  \( AMD = 1 \times 1 \times 11\)} \\ \cline{2-5}
    & R.L. Search& \multicolumn{3}{c|}{ \(NVIDIA \approx 0 + 0 + 5\) ,\quad  \(AMD \approx 0 + 0 + 5\)} \\ \hline
    \end{tabular}
    \vspace{-1.4em}
    \label{tab:exploration_cost}
\end{table}

\section{Related works}
\label{sect:relatedworks}
{
\textbf{GPU accelerated query execution.}
Several open-source GPU-based database systems have been developed, such as Ocelot~\cite{ocelat}, BlazingSQL~\cite{blazingdb}, HeavyDB~\cite{heavydb}, and RAPIDS~\cite{rapids}. A large body of work has extensively explored techniques for optimizing query execution on GPU hardware~\cite{tcuJoin,join1, join2, join2025wu, radixsort, funke18sigmod, jitcompile, vldb2020funke, Damon18Lang, relationaljoin, largescale-tqp}. Some studies have further investigated leveraging PyTorch implementations to accelerate GPU-based query processing~\cite{tensortea_tqp, tqp}. Building upon the significant advancements of prior research, our study addresses the next major bottleneck: data movement to GPUs.}

\noindent
{
\textbf{Optimizing data movement to GPUs.}
Compression has been extensively employed to alleviate data movement bottlenecks to GPUs~\cite{golap, sigmod22shanbhag, hippogriff, vldb10Fang, yinandyang, daemon25nicholson, roller}. Prior studies~\cite{vldb10Fang, hippogriff} propose planners that select optimal combinations of lightweight, column-specific algorithms based on compression ratio and decompression throughput for CPU. Several studies emphasize the importance of designing GPU-friendly decompression algorithm~\cite{afroozeh2024accelerating, sigmod22shanbhag}. More recent work~\cite{golap} explores the use of heavyweight compression algorithms combined with on-the-fly data decompression to enable fine-grained pipelining of data transfers. Our framework further complements these efforts by providing a flexible and extensible infrastructure for developing optimized decompression kernels, thereby supporting a broader spectrum of algorithmic designs and their device scheduling.
Other solutions to manage memory constraints of GPU-based data analytics include hybrid use of CPU and GPU~\cite{pvldb22Yogatama, ghive,Li2025VLDB}, pipelining data movements~\cite{golap, pvldb20Rui}, minimizing redundant movement by lazy scheduling and transfer sharing~\cite{cidr20Raza}, and utilizing all CPU-GPU PCIe links in multi-GPU scenarios~\cite{vortex}.}

\section{Conclusion}
\label{sect:conclusion}
In this paper, we propose a compiler-based framework \proposed to systematically accelerate compressed data movement from CPU to GPU.
This approach mitigates the significant PCIe overhead associated with transferring uncompressed data by leveraging both data-specific characteristics and GPU architectural features.

Both compression ratio and decompression throughput represent critical and interdependent metrics that must be jointly optimized to maximize end-to-end transfer efficiency. While achieving a high compression ratio largely depends on the diversity and adaptability of the available compression algorithm pool, attaining high decompression throughput relies on sophisticated kernel scheduling strategies. \proposed addresses these challenges by introducing three novel design patterns that facilitate flexible optimization across a wide range of compression families on the GPU. This approach not only enables users to efficiently compose custom algorithms to support novel compression strategies and nesting, but it also guarantees their high decompression throughput on heterogeneous devices. Experimental results demonstrate significant advantages over predefined solutions tailored to a limited set of devices.

\bibliographystyle{ACM-Reference-Format}
\bibliography{ref}

\end{document}